# Local Field Statistics in Linear Elastic Unidirectional Fibrous Composites

Tarkes Dora Pallicity[1], Maximilian Krause[2], Thomas Böhlke[2]

[1]Microsolidics Group, Department of Mechanical Engineering, Indian Institute of Technology Guwahati, India

[2]Chair for Continuum Mechanics, Institute of Engineering Mechanics, Karlsruhe Institute of Technology (KIT), Germany
Email addresses: tarkes.pallicity@iitg.ac.in, maximilian.krause@kit.edu, thomas.boehlke@kit.edu.

**Abstract**

Statistical fluctuations of local tensorial fields beyond the mean are relevant to predict localized failure or overall behavior of the inelastic composites. The expression for second moments of the local fields can be established using the Hill-Mandel condition. Complete estimation of statistical fluctuations via second moments is usually ignored despite its significance. In Eshelby-based mean-field approaches, the second moments are evaluated through derivatives of Hill's Polarization tensor ($\mathbb{P}_{\mathrm{o}}$) using a singular approximation. Typically, semi-analytical procedures using numerical integration are used to evaluate the derivatives of the polarization tensor $\mathbb{P}_{\mathrm{o}}$. Here, new analytically derived explicit expressions are presented for calculating the derivatives, specifically for unidirectional fibrous composites with isotropic phases. Full-field homogenization using finite element is used to compute the statistical distribution of local fields (exact solution) for the class of random fibrous microstructures. The mean-field estimates are validated with the exact solution across different fiber volume fractions and aspect ratios. The results indicate that the fiber volume fraction significantly influences the fluctuation of stress tensor invariants, whereas the aspect ratio has minimal effect.

## 1 Introduction

Eshelby based mean-field homogenization approaches have been conventionally used assuming that the stress and strain fields are uniform per phase. They also provide more detailed information at the local scale more than the just average of the fields alone. The objective of this paper is to derive analytical solution of the local field fluctuations of the stress field and its invariants for linear isotropic elastic phases in a unidirectional fibrous composite.

The statistical fluctuation of the stress field can be evaluated through its second-order statistical moments. A full second-order statistical moment is characterized by a positive definite, minor- and major-symmetric fourth-order tensor. Field fluctuations have been incorporated via second moments of the local stress and strain field in second-order homogenization methods (Ponte Castañeda, 2002) to make estimates of the effective nonlinear behavior of heterogenous materials. Second-order methods fundamentally rely on homogenization of linear comparison composite problem. They use specific isotropic projections of the second moments (invariants of the fourth order tensor) as demonstrated for second order models in determining effective viscoelasticity (Badulescu et al., 2015; Lahellec and Suquet, 2007a; Pallicity and Böhlke, 2021), nonlinear viscoelasticity (Kowalczyk-Gajewska et al., 2024), elasto-viscoplasticity (Lahellec and Suquet, 2007b, 2007c) and viscoplastic behavior of composites (Idiart et al., 2006). Evaluation of the full second moments (fourth order tensor) is often ignored in the homogenization problems of fibrous polymer composites.

Full second moments of stress field can be useful in estimating the fluctuation of invariants of the local stress field to predict the localized failure mechanisms. The localized damage in polymers stems from shear yielding (Estevez et al., 2000) and brittle cavitation (Asp et al., 1996). In case of composites, the localized matrix damage lead to failure of composite parts on a structural scale (Talreja, 2006). It has been examined with and without presence of manufacturing defects by Elnekhaily and Talreja, 2023, 2019. These mechanisms are generally triggered by invariants of the state of stress in the matrix phase such as von-Mises stress, hydrostatic pressure and the maximum principal stress.

Expressions to compute full second-moments of local fields in linear problems was first established by (Bobeth and Diener, 1987, 1986). The full second moments for linear homogenization problems in polycrystalline microstructures have been evaluated using a semi-analytical approach (Brenner et al., 2004; Willot et al., 2020). In the case of non-linear (large-strain framework) behavior of phases, an expression for evaluating full-second moments semi-analytically was provided (Das and Ponte Castañeda, 2021; Lopez-Pamies and Castañeda, 2006). A complete analytical solution was derived for calculating second moments in particulate composites with isotropic phases (Krause et al., 2023). However, a comprehensive analytical expression for the direct calculation of second moments in fibrous composites is not yet available.

Full-field homogenization methods based on FFT (Moulinec and Suquet, 1998; Schneider, 2021) and finite elements (FE) provide exact estimate of local distribution of tensorial field quantities. However, full-field methods are computationally intensive. For practical purposes, a statistical estimation of the local fields is often sufficient to compute the effective behavior and investigate localized failure mechanisms in fibrous composite parts. Moreover, implementation of two-scale approaches in structural scale simulations, such as $FE^2$ methods, is computationally demanding. Mean-field methods (Hessman et al., 2021) are an efficient alternative to make a quick close estimate of the mean behavior as well as local field statistics. In view of this, an analytical expression to compute second moments are desirable.

Section 2 and 3 in the current article establish generalized relations for computing second moments of local fields in a small strain framework of a linear elastic homogenization problem consisting of $n$ phases. Sampling of the random data considering normal distribution of local fields is discussed in Section 3.1. A semi-analytical approach is established in Section 4.2 to compute second moments of local field for a two phase composite with ellipsoidal shape of reinforcement and anisotropic phases in the composite. An analytical solution is then derived to compute the same for a composite restricted to ellipsoidal shape of reinforcement and isotropic phases only. Section 5 establishes the framework for full-field simulations using FE method to compute exact statistical distribution of local fields generated in the class of random microstructures. Such nature of microstructures is typically observed after composite fabrication process. The results obtained from both the semi-analytical and analytical solution on distribution of invariants and components of the stress tensor is compared with the solution obtained from the full-field solutions for ellipsoidal shaped fibers reinforced in matrix. The effects of varying the volume fraction and aspect ratio of the reinforcements on the statistical distribution of local fields are also investigated.

Nomenclature

A symbolic tensor notation is used throughout the text. The scalar quantities are denoted by light-face type Latin and Greek characters e.g., $w, K, G$. An orthonormal basis $\{\boldsymbol{e}_1, \boldsymbol{e}_2, \boldsymbol{e}_3\} = \{\boldsymbol{e}_i\}$ of a three-dimensional Euclidean space is used throughout to represent all the tensorial quantities. The first-order tensors are denoted by bold lower-case Latin characters e.g., $\boldsymbol{x}$. The second-order tensors are denoted by bold upper-case Latin characters and Greek characters e.g., $\boldsymbol{I}, \boldsymbol{\Sigma}, \boldsymbol{E}, \boldsymbol{\sigma}, \boldsymbol{\varepsilon}$. The fourth-order tensors are denoted by blackboard bold upper-case Latin characters e.g., $\mathbb{C}, \mathbb{I}$. The major transpose operation is indicated as $(\bullet)^{\mathrm{T_H}}$, left and right minor symmetry is indicated as $(\bullet)^{\mathrm{T_L}}$ and $(\bullet)^{\mathrm{T_R}}$ respectively for fourth order tensors. Similarly, the sixth and eighth order tensor is represented by Fraktur bold upper-case English alphabets e.g. $\boldsymbol{\mathfrak{J}}$ and calligraphic bold upper-case English alphabets e.g. $\boldsymbol{\mathcal{J}}$ respectively. The linear map of a vector over a second order tensor is denoted by $\boldsymbol{Ax} = A_{ij}x_j\boldsymbol{e}_i$, while higher-order linear maps are indicated by $\mathbb{C}[\boldsymbol{\varepsilon}] = C_{ijkl}\varepsilon_{kl}\boldsymbol{e}_i \otimes \boldsymbol{e}_j$. The scalar product between tensors of same order is denoted as $\big((\blacksquare)\cdot(\bullet)\big)_{i_1\dots i_n} = (\blacksquare)_{i_1\dots i_n}(\bullet)_{i_1\dots i_n}$. The tensor product of different orders of tensors are indicated as $\big((\blacksquare)\otimes(\bullet)\big)_{i_1\dots i_m i_n\dots i_o} = (\blacksquare)_{i_1\dots i_m}(\bullet)_{i_n\dots i_o}$. The Rayleigh product with any order of tensor is denoted by $\boldsymbol{R}\star(\bullet) = (\bullet)_{i_1\dots i_n}(\boldsymbol{Re}_{i_1})\otimes\dots\otimes(\boldsymbol{Re}_{i_n})$. If $\boldsymbol{R}$ is a proper orthogonal tensor, it geometrically signifies an active rotation of the tensor, not a passive rotation of the coordinate system. The Kronecker product (Zheng and Spencer, 1993) of any order tensors of is defined as $\big((\blacksquare)\times(\bullet)\big)_{abc\dots ijk\dots mno\dots xyz\dots} = (\blacksquare)_{abc\dots mno\dots}(\bullet)_{ijk\dots xyz\dots}$. The Frobenius norm of a tensor is denoted as for e.g. $\|\bullet\|$. The $n$-times tensor product and Kronecker product of any tensorial quantities with itself is symbolically indicated as $(\bullet)^{\otimes n}$ and $(\bullet)^{\times n}$ respectively. The Kronecker delta is indicated by $\delta_{ij}$. The second, fourth-order (both major and minor symmetric) and eighth order identity tensors is denoted by $\boldsymbol{I}$, $\mathbb{I}^{\mathrm{s}} = \frac{1}{2}(\boldsymbol{I}^{\times 2} + (\boldsymbol{I}^{\times 2})^{\mathrm{T_R}})$ and $\boldsymbol{\mathcal{J}}^{\mathrm{s}}$, respectively. The definition of $\boldsymbol{\mathcal{J}}^{\mathrm{s}}$ is elaborated in Appendix A.1. The fourth-order orthogonal projection tensors are denoted and defined as $\mathbb{P}_1 = \frac{1}{3}(\boldsymbol{I}\otimes\boldsymbol{I})$ and $\mathbb{P}_2 = \mathbb{I}^{\mathrm{s}} - \mathbb{P}_1$ corresponding to the spherical and deviator, respectively. The fourth order positive definite major and minor symmetric elastic stiffness tensor is represented as $\mathbb{C}$. The spatial average of the field quantities over a $\gamma$-phase of a composite is indicated by angular brackets for e.g., $\langle\boldsymbol{\sigma}\rangle_\gamma$.

## 2 Effective Elastic Behavior of Unidirectional Fibrous Composites

Consider the microstructure of a unidirectional (UD) fiber reinforced polymer (FRP) composite occupying a volume $\omega$, consisting of randomly distributed fibers arranged in unidirectional fashion. Under the assumptions of ergodicity and length scale separation, the microstructure of UD FRP composite can be geometrically modelled by a representative volume element (RVE). The domain consists of two phases, with a specific realization for the FRP composite where phase 1 is considered to be matrix. Each $\gamma$-phase is characterized by indicator function $\mathcal{I}_\gamma(\boldsymbol{x})$ (Torquato, 2002) such that,

$$\mathcal{I}_\gamma(\boldsymbol{x}) = \begin{cases} 1 & \boldsymbol{x} \in \omega_\gamma \\ 0 & \text{else} \end{cases}$$

where $\boldsymbol{x}$ is the local coordinates lying in $\omega_\gamma$ of the composite which occupies a volume fraction $c_\gamma$.

### 2.1 Constitutive behavior of the phases in composite

The elastic strain energy density function in a small strain framework for linear elastic phases in an FRP composite can be expressed as

$$w = \frac{1}{2}\mathbb{C} \cdot (\boldsymbol{\varepsilon} \otimes \boldsymbol{\varepsilon}), \tag{1}$$

where $\boldsymbol{\varepsilon}$ is the strain tensor. The state of stress is then given as

$$\boldsymbol{\sigma} = \frac{\partial w}{\partial \boldsymbol{\varepsilon}} = \mathbb{C}[\boldsymbol{\varepsilon}]. \tag{2}$$

If the elastic stiffness tensor $\mathbb{C}$ of the phase is isotropic, it can be defined using the projection tensors as $\mathbb{C} = 3K\mathbb{P}_1 + 2G\mathbb{P}_2$, where $K$ and $G$ are the bulk and shear modulus, respectively. In terms of Young's modulus $E$ and Poisson's ratio $\upsilon$, $K = E/\big(3(1-2\upsilon)\big)$ and $G = E/\big(2(1+\upsilon)\big)$.

### 2.2 Effective behavior of the composite

The homogenized or effective strain energy density function $W$ for the composite, based on the Hill-Mandel condition, is given as

$$W = \frac{1}{2}\overline{\mathbb{C}} \cdot (\boldsymbol{E} \otimes \boldsymbol{E}) = \langle w \rangle = \frac{1}{2}\sum_{\gamma=1}^{n} c_\gamma \mathbb{C}_\gamma \cdot \langle \boldsymbol{\varepsilon}(\boldsymbol{x}) \otimes \boldsymbol{\varepsilon}(\boldsymbol{x}) \rangle_\gamma, \tag{3}$$

where $\mathbb{C}_\gamma$ is the elastic stiffness tensor, which is considered to be homogenous in the $\gamma$-phase of the RVE. The macroscopic terms $\overline{\mathbb{C}}$ and $\boldsymbol{E}$ are the effective stiffness tensor and macroscopic strain tensor, respectively. The state of stress is then given as

$$\boldsymbol{\Sigma} = \frac{\partial W}{\partial \boldsymbol{E}} = \overline{\mathbb{C}}[\boldsymbol{E}], \tag{4}$$

where $\boldsymbol{\Sigma}$ is the effective stress tensor.

Due to the uniqueness of the solution for linear elastic boundary value problems (Jöchen, 2013), a fourth-order strain concentration tensor $\mathbb{A}(\boldsymbol{x})$ linking local and global strains can be defined which is given as,

$$\langle \boldsymbol{\varepsilon}(\boldsymbol{x}) \rangle_\gamma = \mathbb{A}_\gamma[\boldsymbol{E}], \qquad \langle \mathbb{A} \rangle = \mathbb{I}.$$

The effective elastic properties of the composite (Laws 1973) is then given as

$$\overline{\mathbb{C}} = \sum_{\gamma=1}^{n} c_\gamma \mathbb{C}_\gamma \mathbb{A}_\gamma. \tag{5}$$

If the strain concentration tensor is defined then the effective stiffness and the first moments of local fields can be computed, e.g., in the case of Mori-Tanaka method applied to two-phase composites. In other words, in a two-phase composite the localization tensor and the first moments of local fields can be computed if the effective stiffness is known, e.g., using full-field homogenization methods (Levin, 1967).

## 3 Relations for Extracting Local Field Statistics

Consider the microstructure is subjected to homogenous displacement boundary conditions. A small variation of the $\gamma$-phase stiffness $\mathbb{C}_\gamma \rightarrow \mathbb{C}_\gamma + \delta\mathbb{C}_\gamma$ is performed keeping the resulting effective stiffness same (Bobeth and Diener, 1987; Krause et al., 2023). This would cause a variation in the strain field $\boldsymbol{\varepsilon}(\boldsymbol{x}) \rightarrow \boldsymbol{\varepsilon}(\boldsymbol{x}) + \delta\boldsymbol{\varepsilon}(\boldsymbol{x})$. Since the both old and new strain fields have to satisfy the same homogenous displacement boundary condition, hence $\langle\delta\boldsymbol{\varepsilon}(\boldsymbol{x})\rangle = \mathbf{0}$. Hence, the conditions $\langle\delta\boldsymbol{\varepsilon} \cdot \mathbb{C}_\gamma[\boldsymbol{\varepsilon}]\rangle = \langle\delta\boldsymbol{\varepsilon} \cdot \mathbb{C}_\gamma[\delta\boldsymbol{\varepsilon}]\rangle = \langle\delta\boldsymbol{\varepsilon} \cdot \delta\mathbb{C}_\gamma[\delta\boldsymbol{\varepsilon}]\rangle = 0$ must hold true to satisfy the Hill-Mandel condition. Assuming a piecewise uniform stiffness tensor of the phases, the new effective energy can be expressed as

$$
\begin{aligned}
W + \delta W &= \frac{1}{2}\langle(\boldsymbol{\varepsilon} + \delta\boldsymbol{\varepsilon}) \cdot \mathbb{C}[\boldsymbol{\varepsilon} + \delta\boldsymbol{\varepsilon}]\rangle + \frac{c_\gamma}{2}\langle(\boldsymbol{\varepsilon} + \delta\boldsymbol{\varepsilon}) \cdot \delta\mathbb{C}_\gamma(\boldsymbol{\varepsilon} + \delta\boldsymbol{\varepsilon})\rangle_\gamma \\
&= W + \frac{c_\gamma}{2}\langle\boldsymbol{\varepsilon} \cdot \delta\mathbb{C}_\gamma[\boldsymbol{\varepsilon}]\rangle.
\end{aligned}
\tag{6}
$$

The additional term in the effective energy appears due to variation in the energy of the phase with varied stiffness. Using variational derivative together with Equation (6) can be re-written as

$$
\begin{aligned}
\delta W &= \frac{\partial W}{\partial \mathbb{C}_\gamma} \cdot \delta\mathbb{C}_\gamma, \\
\langle\boldsymbol{\varepsilon}(\boldsymbol{x}) \otimes \boldsymbol{\varepsilon}(\boldsymbol{x})\rangle_\gamma &= \frac{2}{c_\gamma}\frac{\partial W}{\partial \mathbb{C}_\gamma}.
\end{aligned}
\tag{7}
$$

In the above equation to compute the second moments of the local strain field, the involved derivatives of the effective energy function still remain unknown. A semi-analytical procedure and analytical solution to the evaluate the derivatives is described in Sect. 4. After Eq. (7) is evaluated, the variance of the strain tensor containing the fluctuations of the local strain field in $\gamma$-phase is then evaluated as

$$\mathbb{K}_{\gamma}^{\varepsilon} = \langle \boldsymbol{\varepsilon} \otimes \boldsymbol{\varepsilon} \rangle_{\gamma} - \langle \boldsymbol{\varepsilon} \rangle_{\gamma} \otimes \langle \boldsymbol{\varepsilon} \rangle_{\gamma}. \tag{8}$$

Similarly, the variance tensors for computing the fluctuations of the local stress field in $\gamma$-phase can be evaluated from the complementary energy function. However, the variance of stress tensor ($\mathbb{K}_{\gamma}^{\sigma}$) can be computed directly from $\mathbb{K}_{\gamma}^{\varepsilon}$ in a linear problem as the phase stiffness $\mathbb{C}_{\gamma}$ is homogeneous over $\gamma$-phase. The relation is given as

$$\mathbb{K}_{\gamma}^{\sigma} = \mathbb{C}_{\gamma} \mathbb{K}_{\gamma}^{\varepsilon} \mathbb{C}_{\gamma} = \langle \boldsymbol{\sigma} \otimes \boldsymbol{\sigma} \rangle_{\gamma} - \langle \boldsymbol{\sigma} \rangle_{\gamma} \otimes \langle \boldsymbol{\sigma} \rangle_{\gamma}. \tag{9}$$

### 3.1 Random sampling and statistics of derived quantities

In this study, analytical approximations are used, which specify only the phase-wise mean and second moments. The uniform spatial distribution of fibers in the microstructures of UD FRP influences the stress at any matrix point due to fiber-matrix interactions and spatial arrangements. Following the central limit theorem, as the RVE size increases, the combined effects of these random stress contributions approach a Gaussian distribution. Consequently, it is assumed that the local field tensorial variables, as well as their linear combinations within each phase, adhere to a Gaussian distribution. In this case, sampling the local field variables, e.g. stress tensor, based on the first and second moments of the stress tensor is possible. This is given as

$$\boldsymbol{\sigma} = \langle \boldsymbol{\sigma} \rangle_{\gamma} + \mathbb{L}[\boldsymbol{\chi}], \tag{10}$$

where $\boldsymbol{\sigma}$ is the sampled stress tensor, $\mathbb{L}$ is the Cholesky decomposition of the positive semi-definite variance tensor $\mathbb{K}^{\sigma}$ such that $\mathbb{K}^{\sigma} = \mathbb{L}\mathbb{L}^{\mathrm{T_H}}$ and $\boldsymbol{\chi}$ is a second order tensor whose components are independent normal random variables with $\langle \boldsymbol{\chi} \rangle_{ij} = 0$ and $\langle \boldsymbol{\chi}^2 \rangle_{ij} = 1$. Invariants of the tensorial field quantities, such as the first principal stress ($\sigma_1$), the equivalent or von-Mises stress $\left(\sigma_{\mathrm{eq}} = \sqrt{\frac{3}{2} \boldsymbol{\sigma}' \cdot \boldsymbol{\sigma}'}\right)$, and the hydrostatic stress ($\sigma_{\mathrm{m}} = \mathrm{tr}(\boldsymbol{\sigma})/3$) of a stress tensor, is assumed to have normal distribution due to the nature of random variable distribution. Based on this, Eq. (10) can be used to sample the invariants of the local field quantities.

The Gaussian probability density of a scalar quantity, e.g. $(\vartheta)$ with normal distribution $\mathcal{N}(\langle\vartheta\rangle, K^{\vartheta})$ or any second order tensor (muti-variate), e.g. $(\boldsymbol{\vartheta})$ with distribution $\mathcal{N}(\langle\boldsymbol{\vartheta}\rangle, \mathbb{K}^{\boldsymbol{\vartheta}})$ can be calculated as

$$\begin{aligned} p(\vartheta) &= \frac{1}{\sqrt{2\pi\sigma}} \exp\left(-\frac{1}{2}(K^{\vartheta})^{-1}(\vartheta - \langle\vartheta\rangle)^2\right), \\ p(\boldsymbol{\vartheta}) &= \frac{1}{\sqrt{(2\pi)^6 \det(\mathbb{K}_\gamma^{\boldsymbol{\vartheta}})}} \exp\left(-\frac{1}{2}(\mathbb{K}_\gamma^{\boldsymbol{\vartheta}})^{-1} \cdot (\boldsymbol{\vartheta} - \langle\boldsymbol{\vartheta}\rangle_\gamma)^{\otimes 2}\right). \end{aligned} \tag{11}$$

## 4 Computation of Derivatives of the Effective Energy with Phase Stiffness

The computation of $\langle \boldsymbol{\varepsilon}(\boldsymbol{x}) \otimes \boldsymbol{\varepsilon}(\boldsymbol{x})\rangle$ in Eq. (7) can be evaluated by taking the anisotropic energy derivatives with phase stiffness $\mathbb{C}_\gamma$. In general, for any material symmetry of $\mathbb{C}_\gamma$ considering Eq. (3) and taking $\partial W/\partial \mathbb{C}_\gamma$ yields

$$\langle \boldsymbol{\varepsilon}(\boldsymbol{x}) \otimes \boldsymbol{\varepsilon}(\boldsymbol{x})\rangle_\gamma = \frac{1}{c_\gamma}\left(\frac{\partial \overline{\mathbb{C}}}{\partial \mathbb{C}_\gamma}\right)^{\mathrm{T}} [\boldsymbol{E} \otimes \boldsymbol{E}], \tag{12}$$

where $\partial\overline{\mathbb{C}}/\partial\mathbb{C}_\gamma$ is an eighth order tensor. In index notation, Eq. (12) can be written as

$$\langle \boldsymbol{\varepsilon}(\boldsymbol{x}) \otimes \boldsymbol{\varepsilon}(\boldsymbol{x})\rangle_{\gamma\, mnop} = \frac{1}{c_\gamma}\left(\frac{\partial \overline{C}_{ijkl}}{\partial C_{\gamma\, mnop}} E_{ij} E_{kl}\right).$$

### 4.1 Two-phase composites with aligned ellipsoids

Consider an RVE of a two-phase composite in which multiple non-overlapping and identical ellipsoidal (prolate spheroid) shaped fibers are embedded in a matrix. The aspect ratio of the ellipsoidal fiber is defined as $a = l/d$, where $l$ and $d$ represents the major and the minor diameter of the fiber, respectively. The major diameter of the ellipsoid is aligned along the $\boldsymbol{e}_3$-direction (indicated in Figure 1 with right-handed co-ordinate system). The spatial arrangement of the ellipsoidal inhomogeneities (described by the two-point correlation function), is assumed to follow a ellipsoidal distribution that is characterized by the shape of the inhomogeneity itself (Castañeda and Willis, 1995; Das and Ponte Castañeda, 2021; Hu and Weng, 2000). Let the compliant phase be the matrix (assigned as $\gamma\ = 1$), characterized by an elastic stiffness tensor $\mathbb{C}_1$ or $\mathbb{C}_\mathrm{m}$ whereas the stiffer reinforcements are characterized by same elastic stiffness tensor $\mathbb{C}_2$ or $\mathbb{C}_\mathrm{f}$. The volume fraction of the matrix is indicated as $c_1$ and the volume fraction of fiber phase is $c_2 = 1 - c_1$.

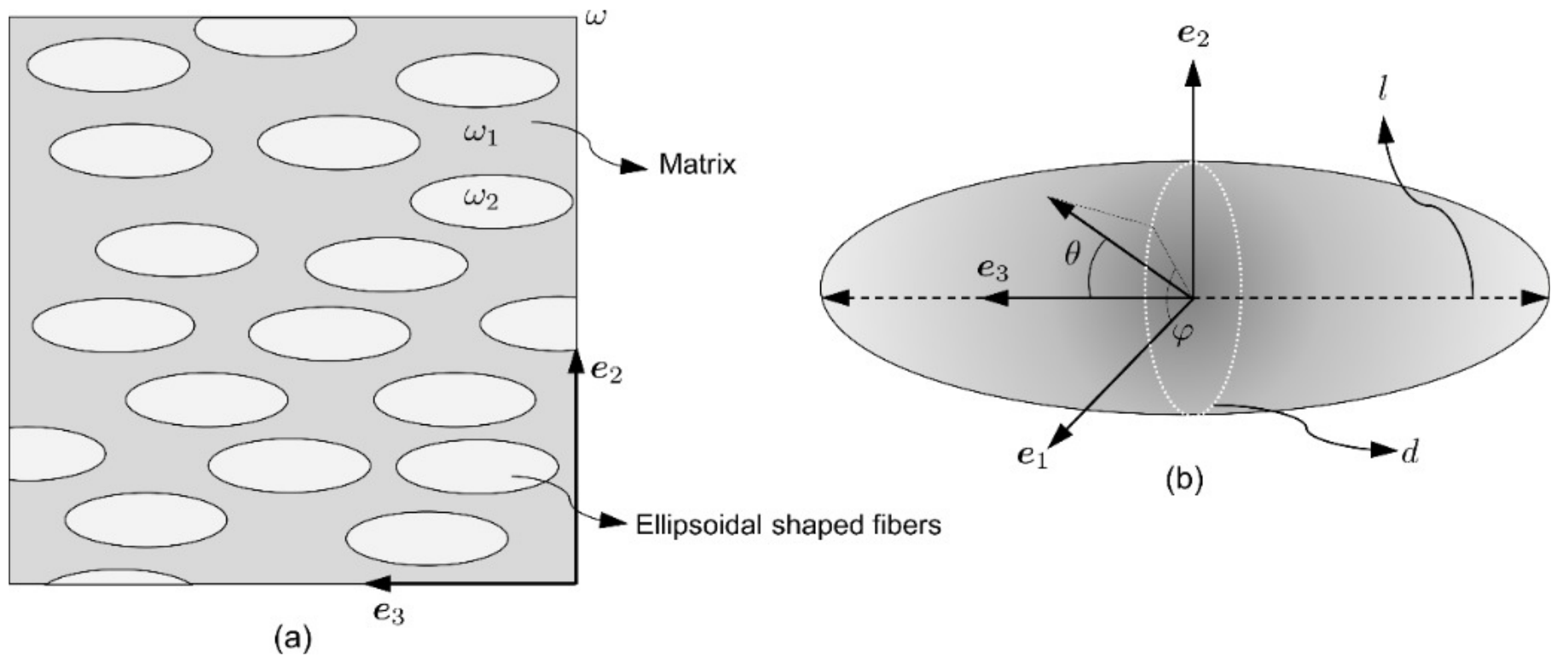

Figure 1 (a) Representative volume element, $\omega$ with random distribution of non-overlapping ellipsoids and oriented along the $\boldsymbol{e}_3$-direction in a matrix. (b) Ellipsoid with major and minor diameter $l$ and $d$, respectively. Aspect ratio is defined as $a = {}^{l}/_{d}$.

### 4.2 Effective stiffness and its derivative: a singular approximation

The effective stiffness and its derivative with respect to phase stiffness are evaluated using a singular approximation, parameterized by reference stiffness $\mathbb{C}_\mathrm{o}$(Fokin, 1974) with formulation considered from Jöchen 2013 . The effective stiffness in Walpole notation (Walpole, 1966) for a singular approximation is expressed as

$$\overline{\mathbb{C}} = \langle \mathbb{L} \rangle^{-1} + \mathbb{C}_\mathrm{o} - \mathbb{P}_\mathrm{o}^{-1}, \qquad \mathbb{L} = (\mathbb{C} - \mathbb{C}_\mathrm{o} + \mathbb{P}_\mathrm{o}^{-1})^{-1}. \tag{13}$$

The Hill polarization tensor $\mathbb{P}_\mathrm{o}$ (major and minor symmetric fourth order tensor) following the formulation by Willis, 1977 is expressed as an integral over the surface of an ellipsoid which is given as

$$\mathbb{P}_\mathrm{o} = \frac{1}{4\pi} \int_S \mathbb{H}(\mathbb{C}_\mathrm{o}, \boldsymbol{n}) \left( \frac{\boldsymbol{n} \cdot (\boldsymbol{A}^{-\mathrm{T}} \boldsymbol{A}^{-1} \boldsymbol{n})}{\det(\boldsymbol{A})^{-\frac{2}{3}}} \right)^{-\frac{3}{2}} \mathrm{d}S, \tag{14}$$

$$\mathbb{H}(\mathbb{C}_\mathrm{o}, \boldsymbol{n}) = \mathbb{I}^\mathrm{s}(\boldsymbol{N} \times \boldsymbol{K}^{-1})\mathbb{I}^\mathrm{s}, \qquad \boldsymbol{K} = \big(\mathbb{I} \times (\boldsymbol{n} \otimes \boldsymbol{n})\big)[\mathbb{C}_\mathrm{o}], \qquad \boldsymbol{A} = \sqrt{\boldsymbol{Z}}.$$

Here, $\boldsymbol{n}$ is the unit normal vector given in a spherical coordinate system i.e., in matrix representation $[\boldsymbol{n}] = [\sin\theta\cos\varphi \quad \sin\theta\sin\varphi \quad \cos\theta]^\mathrm{T}$(see Figure 1). The second order tensor, $\boldsymbol{N} = \boldsymbol{n} \otimes \boldsymbol{n}$ and $\boldsymbol{Z}$ is the ellipsoidal shape of the inhomogeneity. The matrix representation of $\boldsymbol{Z}$ is

$$[\boldsymbol{Z}] = \begin{bmatrix} 1/d & & \\ & 1/d & \\ & & 1/ad \end{bmatrix}.$$

The fourth order symmetric tensor $\mathbb{H}(\mathbb{C}_\mathrm{o}, \boldsymbol{n})$ is expressed in terms of acoustic or Christoffel tensor $\boldsymbol{K}$. Depending on the choice of reference stiffness $\mathbb{C}_\mathrm{o}$, the singular approximation reproduces other results based on the Eshelby solution. In a composite with isotropic phases, the Hashin-Shtrikman bounds (Hashin and Shtrikman, 1962), the Mori-Tanaka method (Mori and Tanaka, 1973) and the singular approximation yield the same effective stiffness. However, Hashin-Shtrikman bounds do not yield the same covariance as compared to the other two methods. For example, in the case of particulate composites, it has been observed that $\mathbb{C}_\mathrm{o} = \mathbb{C}_\mathrm{m}$ results in a better comparisons of mean stiffness and local field statistics (Krause et al., 2023).

Following to Eq. (12), the derivatives of the effective stiffness with phase stiffness for any type of anisotropy can be determined using Eq. $(13)_1$. It is given as

$$
\begin{aligned}
\frac{\partial \overline{\mathbb{C}}}{\partial \mathbb{C}_\gamma} &= \frac{\partial \langle \mathbb{L} \rangle^{-1}}{\partial \mathbb{C}_\gamma} + \frac{\partial \mathbb{C}_\mathrm{o}}{\partial \mathbb{C}_\gamma} - \frac{\partial \mathbb{P}_\mathrm{o}^{-1}}{\partial \mathbb{C}_\mathrm{o}} \, \frac{\partial \mathbb{C}_\mathrm{o}}{\partial \mathbb{C}_\gamma} \\
\Rightarrow \frac{\partial \overline{\mathbb{C}}}{\partial \mathbb{C}_\gamma} &= -(\langle \mathbb{L} \rangle^{-1})^{\times 2} \frac{\partial \langle \mathbb{L} \rangle}{\partial \mathbb{C}_\gamma} + \frac{\partial \mathbb{C}_\mathrm{o}}{\partial \mathbb{C}_\gamma} - (\mathbb{P}_\mathrm{o}^{-1})^{\times 2} \frac{\partial \mathbb{P}_\mathrm{o}}{\partial \mathbb{C}_\mathrm{o}} \, \frac{\partial \mathbb{C}_\mathrm{o}}{\partial \mathbb{C}_\gamma}.
\end{aligned} \tag{15}
$$

Derivative of the terms $\partial \langle \mathbb{L} \rangle / \partial \mathbb{C}_\gamma$ involved in Eq. (15) can be expressed as

$$
\frac{\partial \langle \mathbb{L} \rangle}{\partial \mathbb{C}_\gamma} = -\sum_{\lambda=1}^{2} c_\lambda \mathbb{L}_\lambda^{\times 2} \left( \delta_{\lambda\gamma} \mathbb{I}_8^\mathrm{s} - \frac{\partial \mathbb{C}_\mathrm{o}}{\partial \mathbb{C}_\gamma} - (\mathbb{P}_\mathrm{o}^{-1})^{\times 2} \frac{\partial \mathbb{P}_\mathrm{o}}{\partial \mathbb{C}_\mathrm{o}} \frac{\partial \mathbb{C}_\mathrm{o}}{\partial \mathbb{C}_\gamma} \right), \tag{16}
$$

where $\mathbb{L}_\lambda = (\mathbb{C}_\lambda - \mathbb{C}_\mathrm{o} + \mathbb{P}_\mathrm{o}^{-1})^{-1}$. The objective in Eq. (16) is to evaluate $\partial \mathbb{P}_\mathrm{o} / \partial \mathbb{C}_\mathrm{o}$. These derivatives are evaluated assuming the reference stiffness tensor is anisotropic with major and minor symmetric property.

### 4.2.1 Semi-analytical expression of $\partial \mathbb{P}_\mathrm{o} / \partial \mathbb{C}_\mathrm{o}$ with anisotropic $\mathbb{C}_\mathrm{o}$

Considering the length of the ellipsoids are aligned along $\boldsymbol{e}_3$-direction, the polarization tensor as defined in Eq. (14) can be re-written in index notation as

$$
\begin{aligned}
P^\mathrm{o}_{ijkl} &= \frac{1}{4\pi a^2} \int_S \; c(\theta) H_{ijkl}(\mathbb{C}_\mathrm{o}, \boldsymbol{n}) \mathrm{d}S \;, \\
c(\theta) &= \Big(1 + \big((a^{-2} - 1)\sin^2\theta\big)\Big)^{-3/2}, \\
H_{ijkl} &= \frac{1}{4} (N_{ik}(K_{lj})^{-1} + N_{jk}(K_{li})^{-1} + N_{il}(K_{kj})^{-1} + N_{jl}(K_{ki})^{-1}), \\
K_{lj} &= C^\mathrm{o}_{lmjn} N_{mn}.
\end{aligned} \tag{17}
$$

The term $\partial \mathbb{P}_\mathrm{o} / \partial \mathbb{C}_\mathrm{o}$ is given as

$$
\frac{\partial \mathbb{P}_\mathrm{o}}{\partial \mathbb{C}_\mathrm{o}} = \frac{1}{4\pi a^2} \int_S \; c(\theta) \frac{\partial \mathbb{H}(\mathbb{C}_\mathrm{o}, \boldsymbol{n})}{\partial \mathbb{C}_\mathrm{o}} \mathrm{d}S,
$$

$$
\frac{\partial \mathbb{H}(\mathbb{C}_\mathrm{o}, \boldsymbol{n})}{\partial \mathbb{C}_\mathrm{o}} = \mathbb{I}^\mathrm{s} \left( \frac{\partial (\boldsymbol{N} \times \boldsymbol{K}^{-1})}{\partial \mathbb{C}_\mathrm{o}} \right) \mathbb{I}^\mathrm{s}, \qquad \frac{\partial (\boldsymbol{N} \times \boldsymbol{K}^{-1})}{\partial \mathbb{C}_\mathrm{o}} = -\boldsymbol{N} \times \boldsymbol{K}^{-1} \times \boldsymbol{K}^{-1} \times \boldsymbol{N}.
$$

In index notation,

$$\frac{\partial P^{\mathrm{o}}_{ijkl}}{\partial C^{\mathrm{o}}_{opqr}} = \frac{1}{4\pi a^2}\int_S c(\theta)\,\frac{\partial H_{ijkl}}{\partial C^{\mathrm{o}}_{opqr}}\,\mathrm{d}S,$$

$$\frac{\partial H_{ijkl}}{\partial C^{\mathrm{o}}_{opqr}} = \frac{1}{4}\left(N_{ik}\frac{\partial {K_{lj}}^{-1}}{\partial C^{\mathrm{o}}_{opqr}} + N_{jk}\frac{\partial {K_{li}}^{-1}}{\partial C^{\mathrm{o}}_{opqr}} + N_{il}\frac{\partial {K_{kj}}^{-1}}{\partial C^{\mathrm{o}}_{opqr}} + N_{jl}\frac{\partial {K_{ki}}^{-1}}{\partial C^{\mathrm{o}}_{opqr}}\right), \tag{18}$$

$$\frac{\partial {K_{lj}}^{-1}}{\partial C^{\mathrm{o}}_{opqr}} = -K^{-1}_{lk}\,I^{8s}_{kmsnopqr}\,N_{mn}\,K^{-1}_{sj}.$$

Equation $(18)_1$ can be evaluated using numerical integration methods, employing the analytically derived expression of the integrands. Numerical integration is performed using the adaptive quadrature method in which the step sizes are refined based on error estimates. However, in the isotropic case, an analytical expression for $\partial\mathbb{P}_{\mathrm{o}}/\partial\mathbb{C}_{\mathrm{o}}$ can be derived, as discussed in Sect. 4.2.2. Both approaches are used here to validate the implementation of the derivatives.

In practice, most phases in commercial heterogenous materials, such as polymer composites, are typically isotropic. Consequently, the choice of reference stiffness in mean-field approaches is often isotropic as well. From an implementation standpoint, analytical expressions facilitate faster evaluation of field fluctuations for such cases. However, when dealing with anisotropic reference stiffness and non-ellipsoidal inclusions, numerical integration methods are required.

#### 4.2.2 Analytical solution of $\partial\mathbb{P}_{\mathrm{o}}/\partial\mathbb{C}_{\mathrm{o}}$ with isotropic $\mathbb{C}_{\mathrm{o}}$

The inverse of the acoustic tensor in Eq. $(17)_3$ for isotropic reference stiffness $\mathbb{C}_{\mathrm{o}}$, can be additively split into two components

$$\boldsymbol{K}^{-1} = -\alpha_{\mathrm{o}}\boldsymbol{n}^{\otimes 2} + \beta_{\mathrm{o}}\boldsymbol{I}, \tag{19}$$

where

$$\alpha_{\mathrm{o}} = \frac{3K_{\mathrm{o}} + G_{\mathrm{o}}}{G_{\mathrm{o}}(3K_{\mathrm{o}} + 4G_{\mathrm{o}})}, \qquad \beta_{\mathrm{o}} = \frac{1}{G_{\mathrm{o}}}, \qquad \boldsymbol{n} = \boldsymbol{R}_{\boldsymbol{e}_3}(\varphi)\boldsymbol{R}_{\boldsymbol{e}_2}(\theta)\boldsymbol{e}_3 = \boldsymbol{R}_{\boldsymbol{e}_3}(\varphi)\boldsymbol{n}_{\theta}.$$

The terms $K_\mathrm{o}$ and $G_\mathrm{o}$ are the bulk and shear modulus of isotropic reference stiffness tensor $\mathbb{C}_\mathrm{o}$. The fourth order integrand tensor and the polarization tensor in Eq. (17) can be re-written as:

$$\mathbb{H}(\mathbb{C}_\mathrm{o}, n) = -\alpha_\mathrm{o}\boldsymbol{n}^{\otimes 4} + \beta_\mathrm{o}\mathbb{I}^\mathrm{s}(\boldsymbol{I} \times \boldsymbol{n}^{\otimes 2})\mathbb{I}^\mathrm{s},$$

$$\mathbb{P}_\mathrm{o} = \frac{1}{4\pi a^2}\int_0^{2\pi} \boldsymbol{R}_{\boldsymbol{e}_3} \star \left(-\alpha_\mathrm{o}\int_0^{\pi} f(\theta)\boldsymbol{n}_\theta{}^{\otimes 4}\,\mathrm{d}\theta + \beta_\mathrm{o}\mathbb{I}^\mathrm{s}\left(\boldsymbol{I} \times \int_0^{\pi} f(\theta)\boldsymbol{n}_\theta{}^{\otimes 2}\,\mathrm{d}\theta\right)\mathbb{I}^\mathrm{s}\right)\mathrm{d}\varphi. \tag{20}$$

where $f(\theta) = c(\theta)\sin\theta$ and $a \neq 1$. After substitution and rearranging, the term $\partial\mathbb{H}/\partial\mathbb{C}_\mathrm{o}$ and $\partial\mathbb{P}_\mathrm{o}/\partial\mathbb{C}_\mathrm{o}$from Eq. (18), it can be simplified further to

$$\frac{\partial\mathbb{H}}{\partial\mathbb{C}_\mathrm{o}} = (\mathbb{I}^\mathrm{s})^{\times 2}[\alpha_\mathrm{o}\beta_\mathrm{o}(\boldsymbol{I} \times \boldsymbol{n}^{\otimes 6} + \boldsymbol{n}^{\otimes 6} \times \boldsymbol{I}) + \beta_\mathrm{o}^2\boldsymbol{I} \times \boldsymbol{n}^{\otimes 4} \times \boldsymbol{I} + \alpha_\mathrm{o}^2\boldsymbol{n}^{\otimes 8}],$$

$$\frac{\partial\mathbb{P}_\mathrm{o}}{\partial\mathbb{C}_\mathrm{o}} = \frac{1}{4\pi a^2}(\mathbb{I}^\mathrm{s})^{\times 2}\left[\int_0^{2\pi} \boldsymbol{R}_{\boldsymbol{e}_3} \star \left(\alpha_\mathrm{o}\beta_\mathrm{o}\left(\boldsymbol{I} \times \int_0^{\pi} f(\theta)\boldsymbol{n}_\theta{}^{\otimes 6}\mathrm{d}\theta + \int_0^{\pi} f(\theta)\boldsymbol{n}_\theta{}^{\otimes 6}\mathrm{d}\theta \times \boldsymbol{I}\right) + \beta_\mathrm{o}^2\boldsymbol{I} \times \int_0^{\pi} f(\theta)\boldsymbol{n}_\theta{}^{\otimes 4}\mathrm{d}\theta \times \boldsymbol{I} + \alpha_\mathrm{o}^2\int_0^{\pi} f(\theta)\boldsymbol{n}_\theta{}^{\otimes 8}\mathrm{d}\theta\right)\mathrm{d}\varphi\right]. \tag{21}$$

For completeness, in index notation it is written as

$$\frac{\partial H_{ijkl}}{\partial C^\mathrm{o}_{mnop}} = I^\mathrm{s}_{ijab}I^\mathrm{s}_{klcd}(\alpha_\mathrm{o}\beta_\mathrm{o}((\boldsymbol{n}^{\otimes 6})_{abcmno}\delta_{dp} + (\boldsymbol{n}^{\otimes 6})_{abdmop}\delta_{cn}) + \beta_\mathrm{o}^2(\boldsymbol{n}^{\otimes 4})_{abmo}\delta_{cn}\delta_{dp} + \alpha_\mathrm{o}^2(\boldsymbol{n}^{\otimes 8})_{abcdmnop}), \tag{22}$$

$$
\begin{aligned}
\frac{\partial P^{\mathrm{o}}_{ijkl}}{\partial C^{\mathrm{o}}_{mnop}} = \frac{1}{4\pi a^2} I^{\mathrm{s}}_{ijab} I^{\mathrm{s}}_{klcd} \int_0^{2\pi} \boldsymbol{R}_{\boldsymbol{e}_3} \\
\star \Bigg( \alpha_{\mathrm{o}} \beta_{\mathrm{o}} \Bigg( \left( \left( \int_0^{\pi} f(\theta) \boldsymbol{n}_\theta{}^{\otimes 6} \mathrm{d}\theta \right)_{abcmno} \delta_{dp} \right. \\
+ \left( \int_0^{\pi} f(\theta) \boldsymbol{n}_\theta{}^{\otimes 6} \mathrm{d}\theta \right)_{abdmop} \delta_{cn} \Bigg) \\
+ \beta_{\mathrm{o}}^2 \left( \int_0^{\pi} f(\theta) \boldsymbol{n}_\theta{}^{\otimes 4} \mathrm{d}\theta \right)_{abmo} \delta_{cn} \delta_{dp} \\
+ \alpha_{\mathrm{o}}^2 \left( \int_0^{\pi} f(\theta) \boldsymbol{n}_\theta{}^{\otimes 8} \mathrm{d}\theta \right)_{abcdmnop} \Bigg) \mathrm{d}\varphi.
\end{aligned}
$$

Evaluation of the integral in Eq. $(21)_2$ is the key to the formulation of analytical solution for Eq. (15), Eq. (16) and consequently for Eq. (12) . The integral terms in general form can be written as

$$
\int_0^{2\pi} \boldsymbol{R}_{\boldsymbol{e}_3} \star \int_0^{\pi} f(\theta) \boldsymbol{n}_\theta{}^{\otimes 2l} \mathrm{d}\theta \, \mathrm{d}\varphi, \tag{23}
$$

where $\varphi$-integral projects the results of the $\theta$-integral onto the space of the tensors that are transversely isotropic along the $\boldsymbol{e}_3$-axis. The $\theta$-integral term can be evaluated as

$$
\begin{aligned}
&\int_0^{\pi} f(\theta) \, \boldsymbol{n}_\theta{}^{\otimes 2l} \mathrm{d}\theta \\
&= \begin{cases} 0 \text{ if any of the indices is equal to 2.} \\ 0 \text{ if the number of } 1'\text{s or } 3'\text{s present in the indices is odd.} \\ I_{mn} \quad \text{for all other cases.} \end{cases}
\end{aligned} \tag{24}
$$

where $I_{mn}$ is defined as

$$
I_{mn} = \int_0^{\pi} f(\theta) (\sin\theta)^{2m+1} (\cos\theta)^{2n} \; \mathrm{d}\theta. \tag{25}
$$

The variables $m$ and $n$ are half the number of 1's and 3's present in the indices, respectively. After simplification, (see Appendix A.2) Eq. (25) resolves to a recursive expression

$$
\begin{gathered}
I_{mn} = \sum_{k=0}^{n} \binom{n}{k} (-1)^k I_{(m+k)0}, \\
I_{(m+k)0} = \frac{1}{\gamma}\left(\left(\gamma - 1 - \frac{\gamma+1}{2m}\right) I_{(m+k-1)0} + I_{(m+k-2)0}\right), \\
I_{00} = \frac{2}{\gamma+1}, \qquad I_{10} = \frac{2}{\gamma}\left(\frac{-1}{\gamma+1} + \ g(\gamma)\right),
\end{gathered} \tag{26}
$$

where

$$
\gamma = \left(\frac{1}{a^2} - 1\right) \quad \forall a \neq 1, \qquad g(\gamma) = \begin{cases} \dfrac{\tan^{-1}\sqrt{\gamma}}{\sqrt{\gamma}} & \gamma > 0 \\ \dfrac{\tanh^{-1}\sqrt{-\gamma}}{\sqrt{-\gamma}} & \gamma < 0. \end{cases}
$$

After following the analytical calculations from Eq. (23) to Eq. (26), all the components of the eighth order tensor $\partial \mathbb{P}_o / \partial \mathbb{C}_o$ for a isotropic $\mathbb{C}_o$ can be computed for a composite with ellipsoidal shaped reinforcement. All expressions remain same for a particulate composite with spherical shaped reinforcements ($a = 1$), except the recursive relation in Eq. (26). The term $\partial \mathbb{P}_o / \partial \mathbb{C}_o$ can be recovered same as given in Krause et al. 2023 by setting $\gamma = 0$ and

$$
I_{m0} = \left(\frac{2m}{2m+1}\right) I_{(m-1)0}, \quad I_{00} = 2.
$$

In general, the term $(\partial \mathbb{P}_o / \partial \mathbb{C}_o)_{ijklmnop}$ will have 6561 components. However, the eighth order tensor can be described in a 21×21 symmetric matrix using harmonic decomposition. The space of left, right and main symmetric fourth-order tensors can be decomposed into harmonic subspaces as described by (Forte and Vianello, 1996), each of which is isomorphic to a deviatoric (traceless and fully symmetric) tensor space. If a basis for each of those deviatoric spaces is chosen, a harmonic basis of the original space can be assembled. The fourth order harmonic basis requires five deviatoric basis tensors of second order $\boldsymbol{D}_i^2$ and nine deviatoric basis tensors of fourth order $\mathbb{D}_i^4$, where the $\boldsymbol{e}_3$-axis convention is used (Krause and Böhlke, 2024). The assembly into the original space is done using the isotropic projectors, $\mathbb{P}_1$ and $\mathbb{P}_2$, and the sixth order tensors $\boldsymbol{\mathfrak{J}}_1$ and $\boldsymbol{\mathfrak{J}}_2$, where the orthonormalized definition of $\boldsymbol{\mathfrak{J}}_1$ and $\boldsymbol{\mathfrak{J}}_2$ is used (Krause and Böhlke, 2024). The harmonic basis for left, right and main symmetric fourth-order tensors is given by 21 orthonormal basis tensors $\mathbb{B}_i (i \in [1, 21])$ which is defined as

$$\mathbb{B}_1 = \mathbb{P}_1, \qquad \mathbb{B}_2 = \mathbb{P}_2, \qquad \mathbb{B}_{2+i} = \boldsymbol{\mathfrak{J}}_1[\boldsymbol{D}_i^2], \qquad \mathbb{B}_{7+i} = \boldsymbol{\mathfrak{J}}_2[\boldsymbol{D}_i^2], \qquad \mathbb{B}_{12+j} = \mathbb{D}_j^4,$$

where the index $i \in [1, 5]$ and $j \in [1, 9]$. The definition of the tensors $\boldsymbol{\mathfrak{J}}_1, \boldsymbol{\mathfrak{J}}_2, \boldsymbol{D}_i^2$ and $\mathbb{D}_j^4$ are given in the Appendix A.3. The eighth order tensor $\partial \mathbb{P}_\text{o} / \partial \mathbb{C}_\text{o}$ can be explicitly expressed in terms of these basis tensors as

$$\frac{\partial \mathbb{P}_\text{o}}{\partial \mathbb{C}_\text{o}} = \sum_{\alpha=1}^{21} \sum_{\beta=1}^{21} \frac{\partial P_\alpha}{\partial C_\beta} \mathbb{B}_\alpha \otimes \mathbb{B}_\beta.$$

The term $\partial P_\alpha / \partial C_\beta$ is a 21×21 symmetric matrix i.e., $\partial P_\alpha / \partial C_\beta = \partial P_\beta / \partial C_\alpha$. The non-zero components of $\partial P_\alpha / \partial C_\beta$ is listed in Table 1.

Table 1 Components of $-(\partial P_\alpha/\partial C_\beta)$ indicated as $(\alpha,\beta)$ for isotropic $\mathbb{C}_\mathrm{o}$.

| | |
|---|---|
| (1,1) | $I_{00}\,(\zeta+1)^2/18\eta^2$ |
| (2,1) | $\sqrt{5}I_{00}\,(\zeta+1)^2/45\eta^2$ |
| (2,2) | $I_{00}\,(8(\zeta+1)^2+9)/180\eta^2$ |
| (7,1) | $(2I_{00}-3I_{10})(\zeta+1)^2/18\eta^2$ |
| (7,2) | $2\sqrt{5}\left(\dfrac{\partial P_7}{\partial C_1}\right)/5$ |
| (3,3); (4,4) | $(2I_{20}\zeta^2+(2I_{10}+3I_{20})\zeta+2I_{10}+I_{20})/24\eta^2$ |
| (5,5); (6,6) | $\begin{pmatrix}8(I_{10}-I_{20})\zeta^2+(2I_{00}+11I_{10}-12I_{20})\zeta+\\ 2I_{00}+3I_{10}-4I_{20}\end{pmatrix}/24\eta^2$ |
| (7,7) | $\begin{pmatrix}2(4I_{00}-12I_{10}+9I_{20})\zeta^2+(16I_{00}-39I_{10}+27I_{20})\zeta\\ +8I_{00}-15I_{10}+9I_{20}\end{pmatrix}/36\eta^2$ |
| (12,1) | $\sqrt{14}\left(\dfrac{\partial P_7}{\partial C_1}\right)/7$ |
| (12,2) | $\sqrt{70}\begin{pmatrix}16(2I_{00}-3I_{10})\zeta^2+32(2I_{00}-3I_{10})\zeta+\\ 25(2I_{00}-3I_{10})\end{pmatrix}/5040\eta^2$ |
| (8,3); (9,4) | $\sqrt{14}\,(2I_{20}\zeta^2+(2I_{10}+3I_{20})\zeta+2I_{10}+I_{20})/168\eta^2$ |
| (10,5); (11,6) | $\sqrt{14}\begin{pmatrix}8(I_{10}-I_{20})\zeta^2+(2I_{00}+11I_{10}-12I_{20})\zeta+\\ 2I_{00}+3I_{10}-4I_{20}\end{pmatrix}/168\eta^2$ |
| (12,7) | $\sqrt{14}\begin{pmatrix}(8I_{00}-24I_{10}+18I_{20})\zeta^2+\\ (16I_{00}-39I_{10}+27I_{20})\zeta+8I_{00}-15I_{10}+9I_{20}\end{pmatrix}/252\eta^2$ |
| (8,8); (9,9) | $(16I_{20}\zeta^2+(16I_{10}+24I_{20})\zeta+36I_{00}-20I_{10}+17I_{20})/672\eta^2$ |
| (10,10); (11,11) | $\begin{pmatrix}32(I_{10}-I_{20})\zeta^2+4(2I_{00}+11I_{10}-12I_{20})\zeta\\ +8I_{00}+39I_{10}-34I_{20}\end{pmatrix}/336\eta^2$ |

Table 1 (Contd.) Components of $-\partial P_\alpha/\partial C_\beta$ indicated as $(\alpha,\beta)$ for isotropic $\mathbb{C}_o$.

| | |
|---|---|
| (12,12) | $\left(\begin{matrix}16(4I_{00}-12I_{10}+9I_{20})\zeta^2+\\8(16I_{00}-39I_{10}+27I_{20})\zeta\\+82I_{00}-174I_{10}+153I_{20}\end{matrix}\right)\Big/1008\eta^2$ |
| (21,1) | $\sqrt{70}\,((8I_{00}-40I_{10}+35I_{20})(\zeta+1)^2)/840\eta^2$ |
| (21,2) | $\sqrt{14}\,(8I_{00}-40I_{10}+35I_{20})(4(\zeta+1)^2-3)/1680\eta^2$ |
| (17,3); (18,4) | $\sqrt{42}\left(\begin{matrix}2(6I_{20}-7I_{30})\zeta^2+(12I_{10}-3I_{20}-14I_{30})\zeta\\+12I_{10}-15I_{20}\end{matrix}\right)\Big/336\eta^2$ |
| (19,5); (20,6) | $\sqrt{21}\left(\begin{matrix}8(4I_{10}-11I_{20}+7I_{30})\zeta^2+\\(8I_{00}+16I_{10}-83I_{20}+56I_{30})\zeta\\+(8I_{00}-16I_{10}+5I_{20})\end{matrix}\right)\Big/336\eta^2$ |
| (21,7) | $\sqrt{70}\left(\begin{matrix}(16I_{00}-104I_{10}+190I_{20}-105I_{30})\zeta^2+\\(32I_{00}-148I_{10}+215I_{20}-105I_{30})\zeta+\\(16I_{00}-44I_{10}+25I_{20})\end{matrix}\right)\Big/840\eta^2$ |
| (17,8); (18,9) | $\sqrt{3}\left(\begin{matrix}(24I_{20}-28I_{30})\zeta^2+(24I_{10}-6I_{20}-28I_{30})\zeta\\+(12I_{00}-30I_{10}+15I_{20})\end{matrix}\right)\Big/336\eta^2$ |
| (19,10); (20,11) | $\sqrt{6}\left(\begin{matrix}16(4I_{10}-11I_{20}+7I_{30})\zeta^2+\\2(8I_{00}+16I_{10}-83I_{20}+56I_{30})\zeta\\+16I_{00}-20I_{10}-5I_{20}\end{matrix}\right)\Big/672\eta^2$ |
| (21,12) | $\sqrt{5}\left(\begin{matrix}(32I_{00}-208I_{10}+380I_{20}-210I_{30})\zeta^2+\\(64I_{00}-296I_{10}+430I_{20}-210I_{30})\zeta\\+20I_{00}-10I_{10}-25I_{20}\end{matrix}\right)\Big/840\eta^2$ |
| (13,13); (14,14) | $(I_{40}\zeta^2+4I_{30}\zeta+4I_{20})/32\eta^2$ |
| (15,15); (16,16) | $(8(I_{30}-I_{40})\zeta^2+(18I_{20}-17I_{30})\zeta+8I_{10}-6I_{20})/32\eta^2$ |
| (17,17); (18,18) | $\left(\begin{matrix}2(36I_{20}-84I_{30}+49I_{40})\zeta^2+(72I_{10}-144I_{20}+77I_{30})\zeta\\+8I_{00}+8I_{10}-11I_{20}\end{matrix}\right)\Big/112\eta^2$ |
| (19,19); (20,20) | $\left(\begin{matrix}(128I_{10}-576I_{20}+840I_{30}-392I_{40})\zeta^2+\\(32I_{00}-48I_{10}+18I_{20}+7I_{30})\zeta\\+32I_{00}-40I_{10}+18I_{20}\end{matrix}\right)\Big/224\eta^2$ |
| (21,21) | $\left(\begin{matrix}(64I_{00}-640I_{10}+2160I_{20}-2800I_{30}+1225I_{40})\zeta^2+\\(128I_{00}-480I_{10}+720I_{20}-350I_{30})\zeta\\+96I_{00}-160I_{10}+90I_{20}\end{matrix}\right)\Big/560\eta^2$ |

$\zeta=-(3K_o+G_o)/(3K_o+4G_o)$, $\eta=Ga$. See Eq. (25) and (26) for definition of $I_{mn}$.

## 5 Statistics of Local Stress Field in Fibrous Composite

The exact statistical distribution of local fields is obtained through full-field analysis, which is used to validate the semi-analytical (S-A) and analytical solution (A) discussed in Sect. 4.2.1 and Sect. 4.2.2, respectively. In this study, full-field analysis of microstructures is performed using FE for uniform distribution of non-overlapping fibers (inhomogeneities) with a UD arrangement within a matrix. The fiber volume fraction considered in this study is $c_\mathrm{f} < 0.25$ which is commonly noticed in automotive applications (Görthofer et al., 2019) and $a < 5$ to reduce the geometrical complexity and computational time. In view of this, a random sequence adsorption algorithm is sufficient to generate artificial microstructures for full-field analysis. This approach is implemented in an in-house developed code to generate random position vectors of inhomogeneities considering the constraints of periodicity and non-overlap of fibers. Thanks to the open-source tool Ellipsoid toolbox (Gagarinov and Kurzhanskiy, 2014), which is used to detect non-overlap of the ellipsoids (fiber) during the generation of random position vectors within a constrained space of 3-D cuboid (matrix). The geometrical model generation of the artificial microstructures using these position vectors and discretizing with 10-node quadratic tetrahedron elements are seamlessly done using NETGEN (Schöberl, 1997). An in-house code is used to apply periodic boundary condition on the free surfaces of the RVE (Albiez et al., 2019). The traction and displacements are considered to be continuous across the interface of the matrix and inhomogeneities. Numerical analysis of the UD FRP microstructures, including long- and short- fiber reinforced polymer (termed as LFRP and SFRP, respectively), is performed in a commercial FE platform (ABAQUS).

In this work, a glass fiber-reinforced polymer is considered due to its relatively high elastic contrast of 21.5. The matrix material is elastic unsaturated polyester–polyurethane hybrid (UPPH) with properties Young's modulus, $E = 3.4$ GPa and Poisson's ratio, $\nu = 0.385$. Experimentally characterized properties of the polymer matrix are used (Kehrer, 2019; Trauth and Weidenmann, 2018). The glass fibers with elastic properties $E = 73$ GPa and $\nu = 0.22$ are ellipsoidal in shape. The macroscopic strain tensor considered in the full-field simulations is a combination of a uniaxial extension ($E_{22}$) normal to the fiber direction and equal shear on the planes parallel to the fiber direction ($E_{12}, E_{23}$) which is given as

$$\boldsymbol{E} = 0.005(\boldsymbol{e}_2 \otimes \boldsymbol{e}_2 + \boldsymbol{e}_1 \otimes^{\mathrm{s}} \boldsymbol{e}_2 + \boldsymbol{e}_2 \otimes^{\mathrm{s}} \boldsymbol{e}_3). \tag{27}$$

### 5.1 Short Fiber Reinforced Polymer (SFRP) Composites

The solution for $\partial\mathbb{P}_{\mathrm{o}}/\partial\mathbb{C}_{\mathrm{o}}$ for fiber volume fraction$c_{\mathrm{f}} = 0.25$ and an aspect ratio $a = 5$ (ellipsoidal shaped fibers) is obtained using semi-analytical procedure and analytical solution for isotropic reference stiffness (see Table 1). Estimation of statistics of invariants of the local stress field as well as components of stress tensor is obtained from solution $\partial\mathbb{P}_{\mathrm{o}}/\partial\mathbb{C}_{\mathrm{o}}$. Full-field simulations for SFRP microstructures with 25 ellipsoidal shaped particles in RVE is carried out using FEM.

Figure 2(a) shows the statistical distribution of the von-Mises stress ($\sigma^{\mathrm{m}}_{\mathrm{mises}}$), first principal stress ($\sigma^{\mathrm{m}}_{1}$) and hydrostatic stress ($\sigma^{\mathrm{m}}_{\mathrm{o}}$) in the matrix domain obtained using three different approaches. All three invariants are calculated from the sampled stress tensor data using Eq. (8) to (12) via the solution of $\partial\mathbb{P}_{\mathrm{o}}/\partial\mathbb{C}_{\mathrm{o}}$. This is because the mean and variance of the stress tensor invariants are in general inaccessible through the mean-field methods when the fluctuations are non-zero. The invariants exhibit nonlinearity in terms of the tensor components due to the presence of squared terms, as well as quadratic and cubic roots. In contrast, hydrostatic stress is a linear function, as it is calculated as the mean of the normal stress components.

Figure 2(b) shows the sampled components of the stress tensor $\sigma_{22}$ and $\sigma_{23}$, compared with the exact distribution (histogram) and the fitted distribution. In both plots, the computed mean of the stress components and invariants closely aligns with the fitted normal distribution obtained from the full-field simulation. The computed solution captures the overall trend, though not precise when compared with the exact distribution. The full-field discrete results include higher-order statistical moments, which prevent the distribution from being perfectly described by a Gaussian function.

The effect of fiber aspect ratio is investigated by keeping the fiber volume fraction fixed. Full-field simulations are performed with 25 ellipsoidal shaped particles having an aspect ratio $a = 3$ in the RVE. Figure 3(a) and (b) show the statistical distribution of the $\sigma^{\mathrm{m}}_{\mathrm{mises}}$, $\sigma^{\mathrm{m}}_{1}$ and $\sigma^{\mathrm{m}}_{\mathrm{o}}$, and component of stress tensor $\sigma_{22}$ and $\sigma_{23}$, respectively in the matrix domain. The overall trend of the exact distribution is well captured by the solution obtained using mean-field approaches.

The standard deviation (*SD*) of invariants (in MPa) computed from the fitted distribution is compared with the *SD* of the invariants using sampled data for different aspect ratio, as shown in Figure 4. The fluctuations of the invariants obtained from semi-analytical solution and analytical solution is identical for any aspect ratio. The difference relative to the exact solution is found to be less than 3 MPa. Additionally, the *SD* of the $\sigma_{22}$ and $\varepsilon_{22}$ obtained from the full-field solution for both compliant matrix and stiffer reinforcement phase is plotted in Figure 5. It is observed that the fluctuations of the deformation field are significantly weaker in the fiber phase compared to the matrix phase. This is consistent with the fact that the stiffness of the inhomogeneity is relatively high (with an elastic modulus contrast of approximately 25), compared to the matrix, leading to almost homogenous deformations. As an approximation, fluctuation of the deformation field in the inhomogeneity can be neglected, aligning with the assumption in Eshelby-based homogenization where the fluctuation field is considered as zero within the inclusion. Using singular approximation, the term $\partial \mathbb{P}_{\mathrm{o}} / \partial \mathbb{C}_{\mathrm{f}}$ is an eighth order zero tensor ($\boldsymbol{\mathcal{O}}$) considering $\mathbb{C}_{\mathrm{o}} = \mathbb{C}_{\mathrm{m}}$. Upon, further simplification, one obtains the fluctuations of the strain field in the fiber to be zero i.e., $\mathbb{K}_{\mathrm{f}}^{\varepsilon} = \mathbb{O}$, and hence $\langle \boldsymbol{\varepsilon} \otimes \boldsymbol{\varepsilon} \rangle_{\mathrm{f}} = \langle \boldsymbol{\varepsilon} \rangle_{\mathrm{f}} \otimes \langle \boldsymbol{\varepsilon} \rangle_{\mathrm{f}}$. This has been proven analytically in (Das and Ponte Castañeda, 2021; Krause et al., 2023). Here it is confirmed semi-analytically and analytically in which the strain covariance tensor is obtained to be zero. In contrast, stress field fluctuations are comparable in both phases as observed in full-field simulations, even though the constitutive relation is linear. However, the semi-analytical solution still predicts zero fluctuations in stress field within the stiffer phase due to the linear constitutive relations.

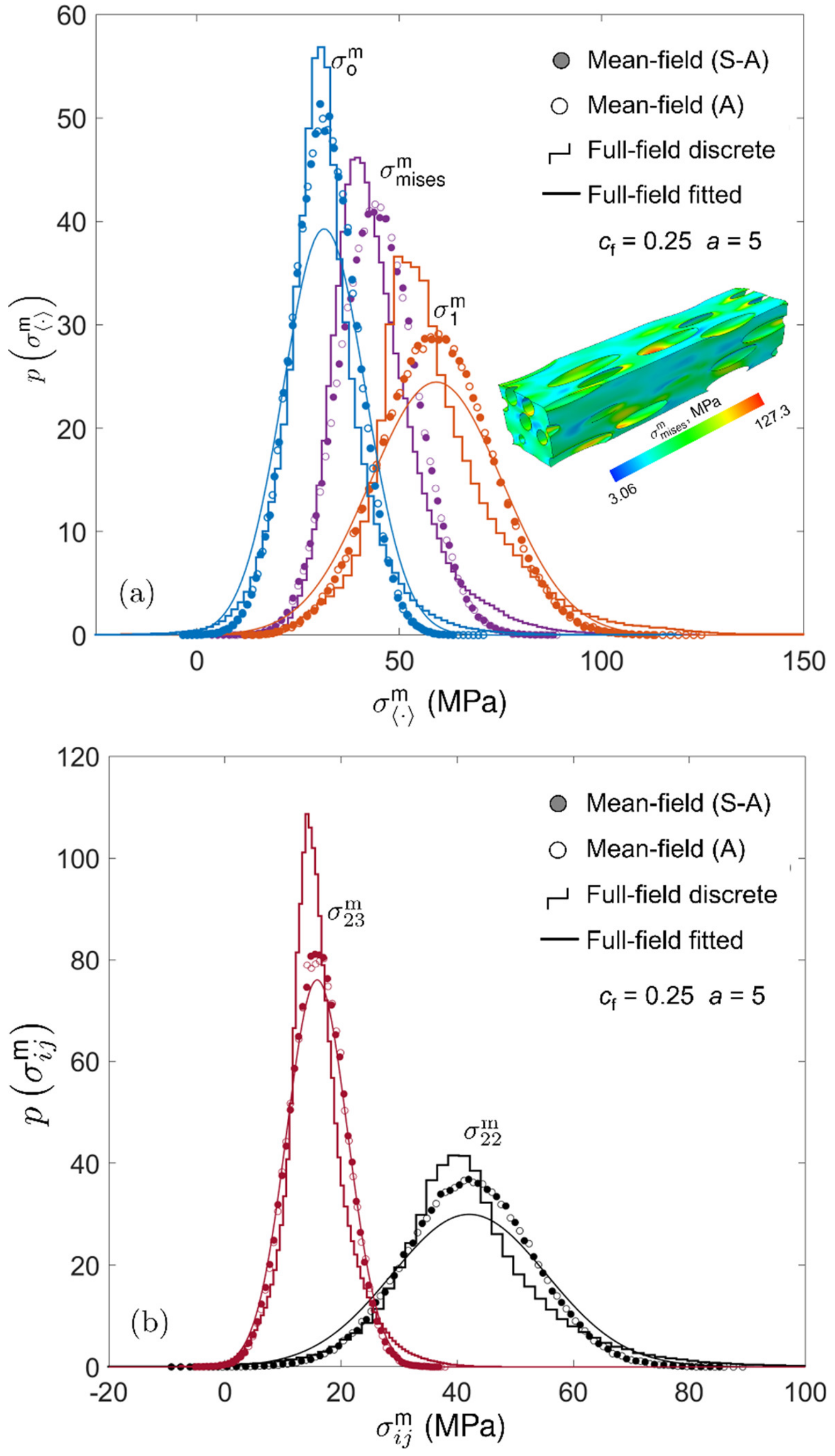

Figure 2 Statistical distribution of local fields in the matrix domain of unidirectional short fiber reinforced composite with 25% fiber volume fraction and aspect ratio of ellipsoid is 5 (a) Stress Invariants (b) Components of stress tensor. Mean-field (S-A) and Mean-field (A) indicates the statistical distribution obtained using semi-analytical and analytical approaches. Actual data obtained from full-field simulations are plotted as outlined histogram plot. The fitted distribution is indicated as solid line.

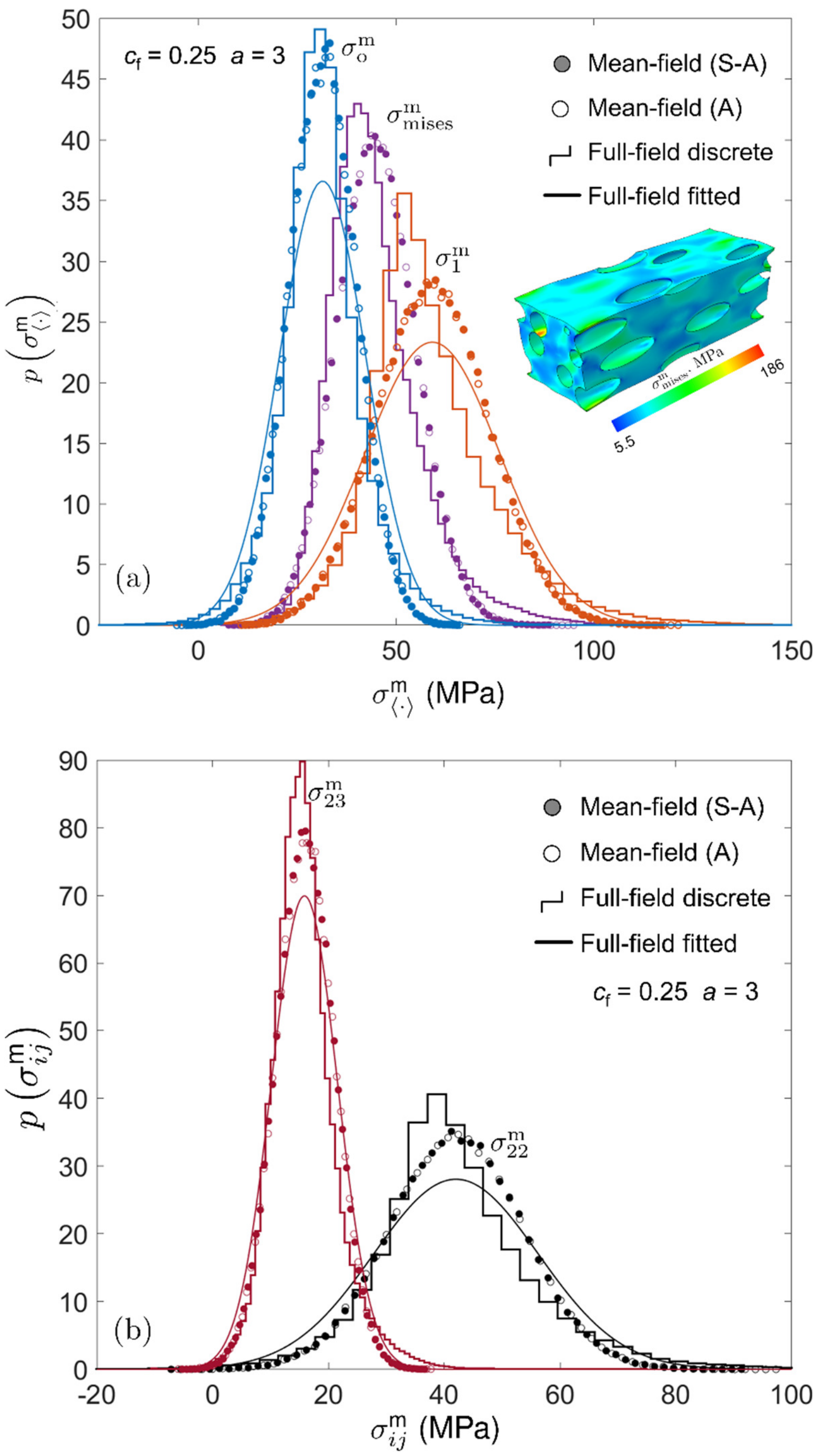

Figure 3 Statistical distribution of local fields in the matrix domain of unidirectional short fiber reinforced polymer composite with 25% fiber volume fraction and aspect ratio is 3 (a) Invariants (b) Components of stress tensor. Mean-field (S-A) and Mean-field (A) indicates the distribution obtained using semi-analytical and analytical approaches. The actual data from full-field simulations are represented by outlined histogram plot. The fitted distribution to the actual data is indicated as a solid line.

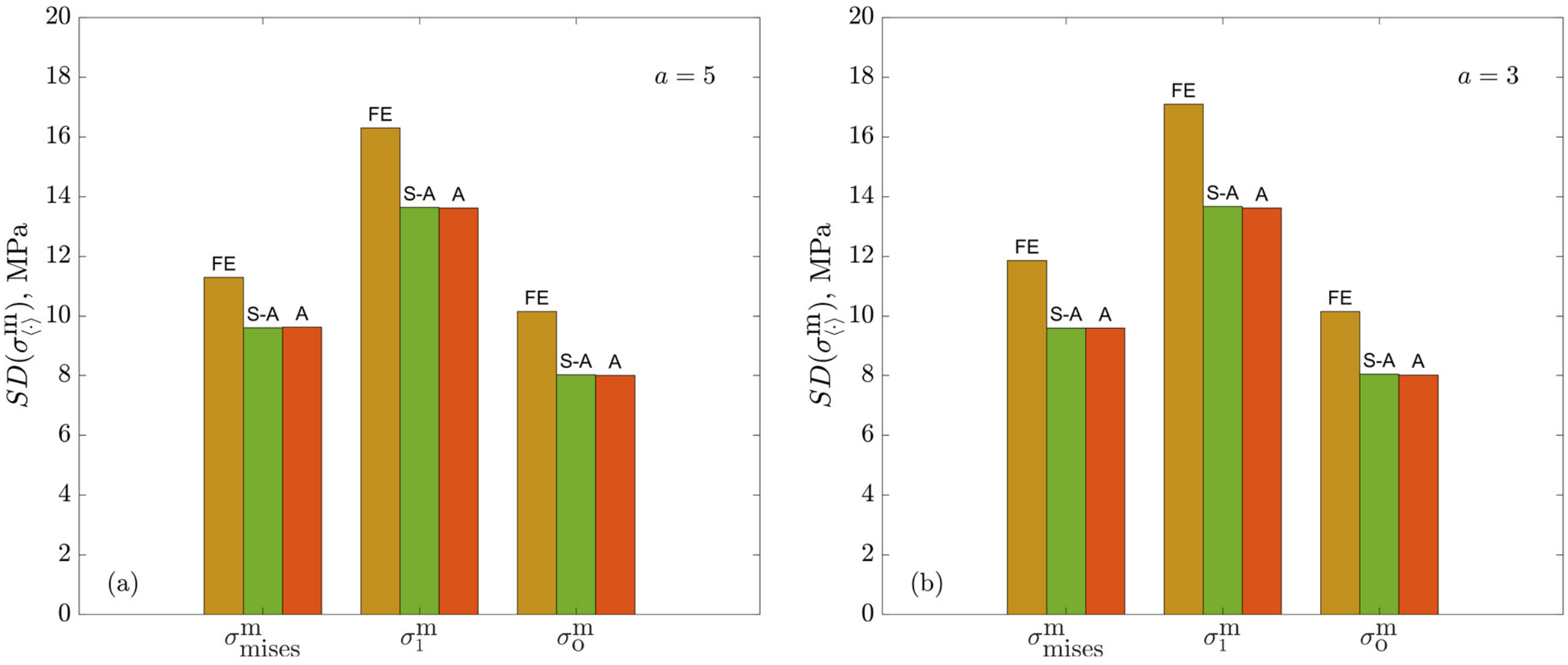

Figure 4 Variance of the invariants of stress tensor in the matrix domain of the ellipsoidal shaped fibers reinforced polymer composite with fiber volume fraction of 25% (a) aspect ratio of ellipsoidal short fiber, $a = 5$ and (b) $a = 3$. Solution is computed using full-field (FE), semi-analytical (S-A) and the analytical (A) approach.

## 5.2 Long Fiber Reinforced Polymer (LFRP) Composites

The term $\partial\mathbb{P}_{\mathrm{o}}/\partial\mathbb{C}_{\mathrm{o}}$ for inhomogeneities with $a = 1000$ (cylindrical shaped long fibers) and $c_{\mathrm{f}} = 0.15$ is solved using analytical solution as well the semi-analytical approach for isotropic reference stiffness. Full-field simulations for LFRP microstructures with 60 cylindrical shaped particles are carried out using FEM.

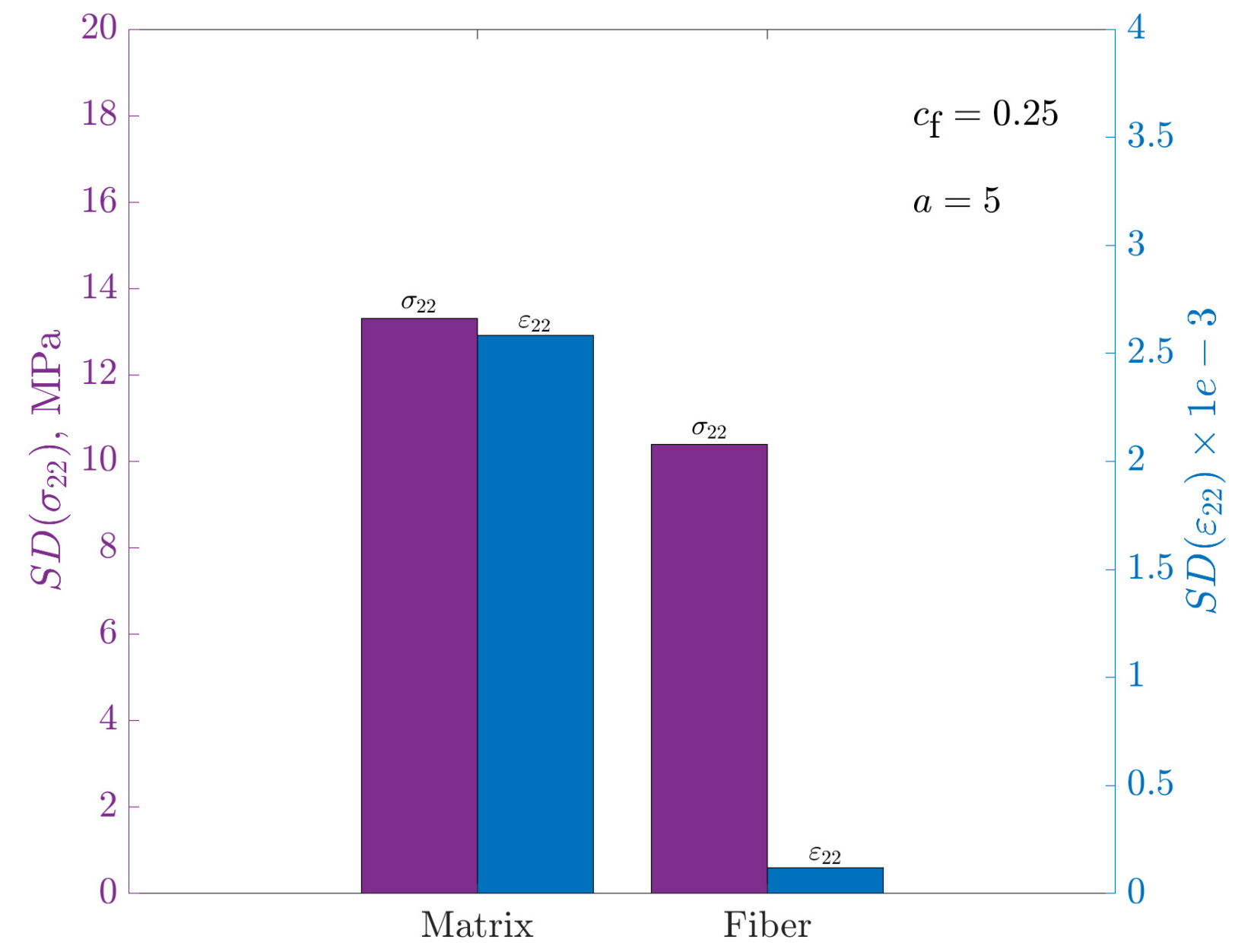

Figure 5 Variance of local $\sigma_{22}$ and $\varepsilon_{22}$ field in fibrous composite with ellipsoidal shaped inhomogeneities of aspect ratio $a = 5$ and volume fraction $c_f = 0.25$ computed using full-field method. Fluctuations of deformations in the stiffer reinforcements are very weak relative to compliant phase. Fluctuations of stress fields are comparable.

Figure 6(a) shows the statistical distribution of $\sigma^{m}_{mises}$, $\sigma^{m}_{1}$ and $\sigma^{m}_{o}$ in the matrix domain obtained using three different approaches. Figure 6(b) shows the sampled components of the stress tensor $\sigma_{22}$ and $\sigma_{23}$, compared with the exact distribution (histogram) and the fitted distribution. In both plots, the computed mean of the stress components and invariants closely aligns with the fitted normal distribution obtained from the full-field simulation. The effect of volume fraction on predicting local field statistics in LFRP is investigated for $c_f = 0.25$. Full-field simulations are performed considering 20 cylinders in the RVE. Figure 7 (a) and (b) shows the statistical distribution of the $\sigma^{m}_{mises}$, $\sigma^{m}_{1}$ and $\sigma^{m}_{o}$, and stress tensor components $\sigma_{22}$ and $\sigma_{23}$, respectively in the matrix domain, as obtained using semi-analytical, analytical and FE simulations. The sampled data from mean-field approach matches well with the fitted distribution obtained from the full-field solution.

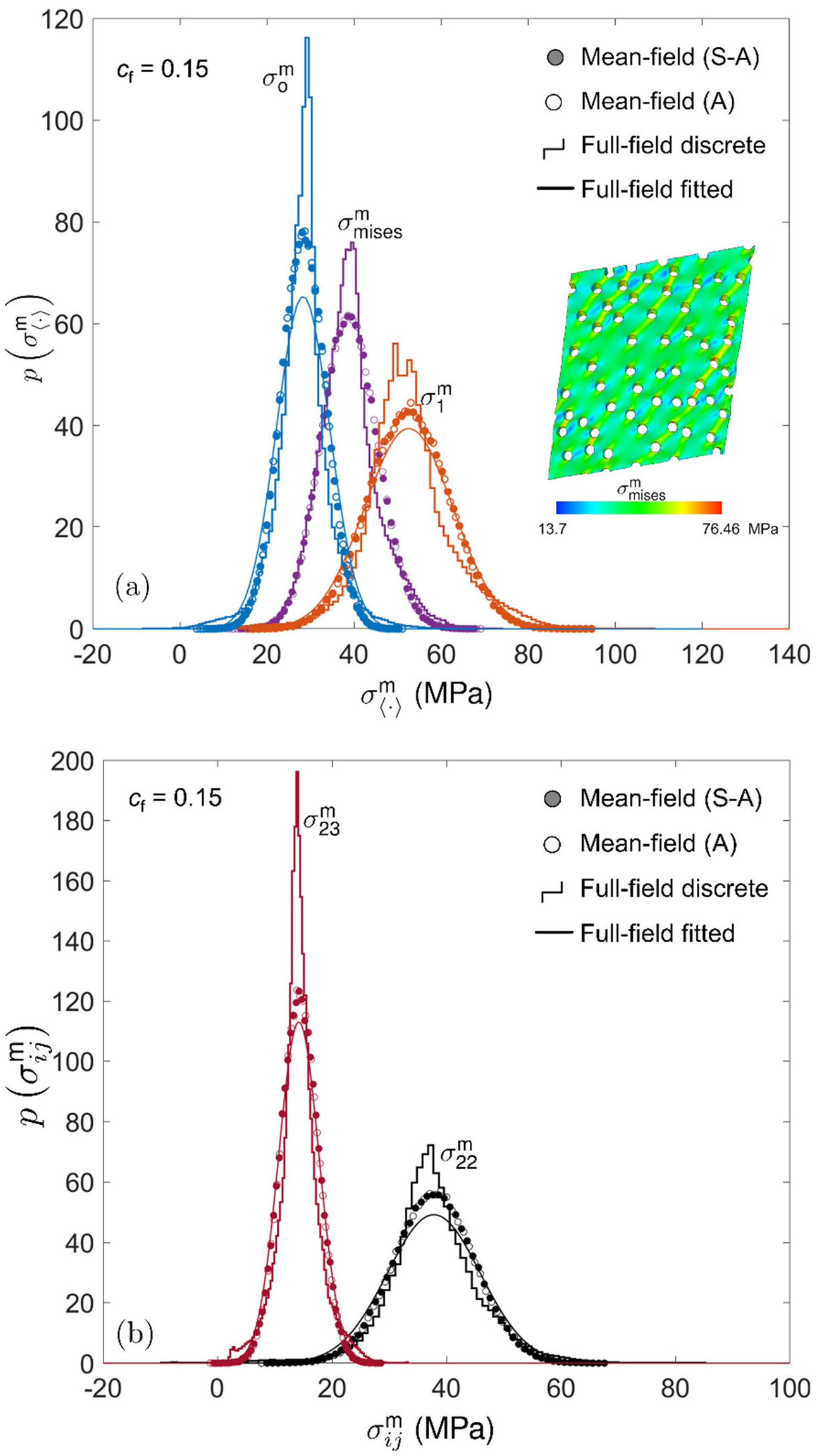

Figure 6 Statistical distribution of local fields in the matrix domain of the unidirectional long fiber reinforced polymer composite with 15% fiber volume fraction (a) Stress invariants (b) Components of the stress tensor. MFH (S-A) and MFH (A) indicates the statistical distribution obtained using semi-analytical and analytical approaches, respectively. Actual data obtained from full-field simulations are represented by the outlined histogram plot, while the fitted distribution is indicated as a solid line.

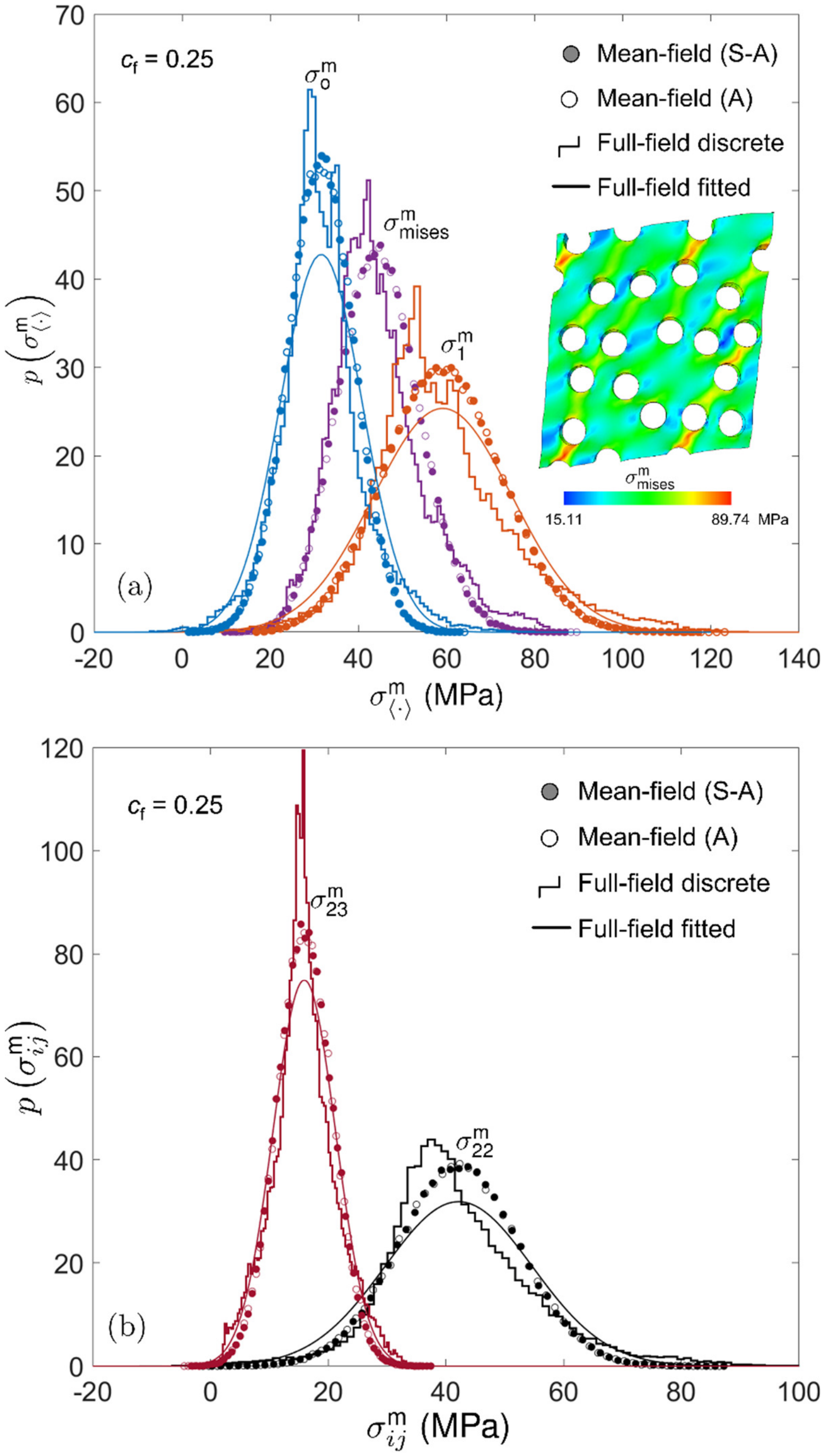

Figure 7 Statistical distribution of the local fields in the matrix domain of unidirectional long fiber reinforced polymer composite with 25% fiber volume fraction (a) Invariants of stress tensor (b) Components of stress tensor. MFH (S-A) and MFH (A) indicates the statistical distribution obtained using semi-analytical and analytical approaches. The actual data from full-field simulations are represented by outlined histogram plot. The fitted distribution from the full-field data is indicated as a solid line.

The standard deviation ($SD$ in MPa) computed from the fitted distribution is compared with the $SD$ of the invariants using sampled data for different volume fractions as shown in Figure 8. The $SD$ of the invariants obtained from analytical solution and semi-analytical solution is identical. However, the difference in the $SD$ is less than 2.5 MPa when compared with the fitted distribution of full-field data. Overall, the mean-field approaches effectively capture the trend of the exact distribution. A comparison between Figure 8(a) and (b) reveals that the fluctuation increases with higher volume fraction of reinforcement. This is because the particle interaction is intensified as the volume fraction increases, thereby leading to higher fluctuations. Comparing the effect of aspect ratio on fluctuation of invariants from Figure 4 and Figure 8(b) (aspect ratio is 3 , 5 and 1000 for fixed $c_{\mathrm{f}} = 0.25$ ), it is clearly evident that the fluctuation is nearly same when $a > 1$. This observation is further supported by visual comparison of Figure 2, Figure 3 and Figure 7 for different aspect ratio with volume fraction of $c_{\mathrm{f}} = 0.25$.

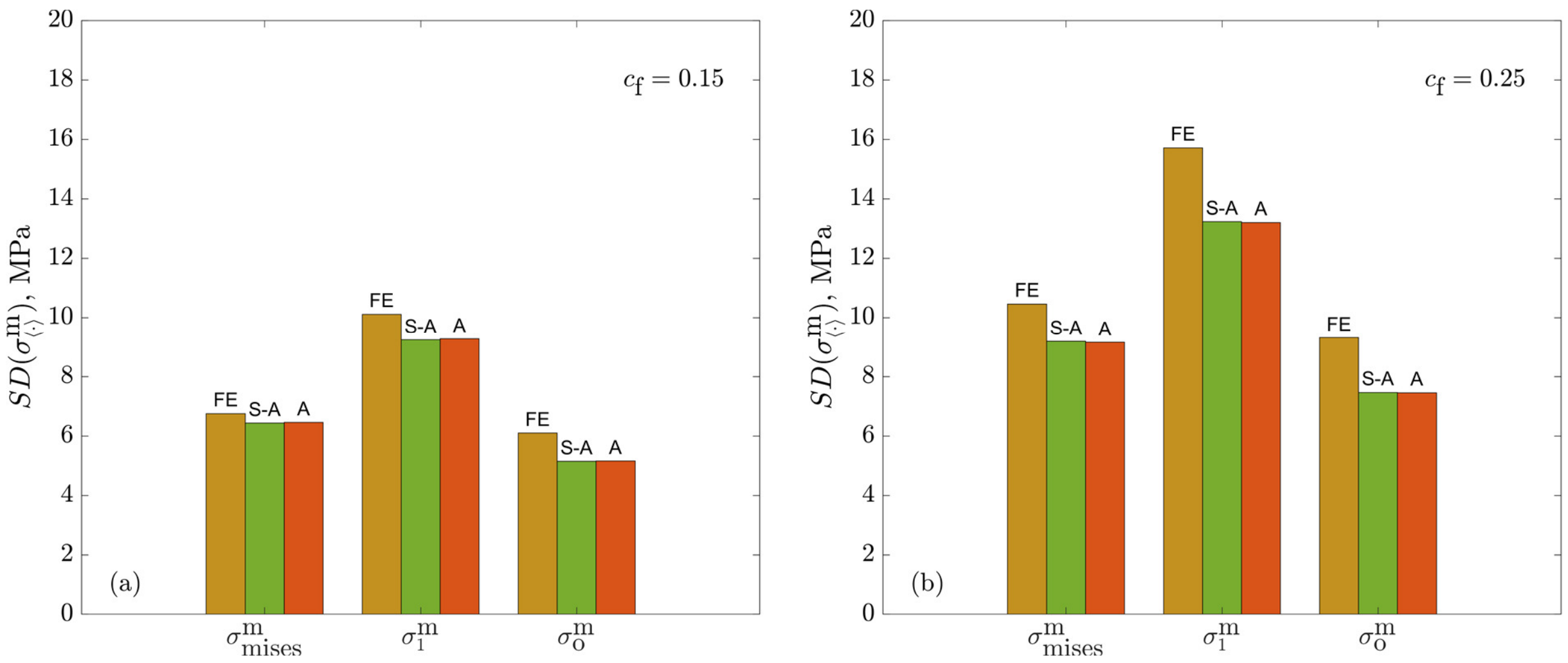

Figure 8 Variance of the invariants of stress tensor in the matrix domain of the long fiber reinforced polymer composite, computed using full-field (FE), semi-analytical (S-A) and the analytical approach (A). (a) Volume fraction of fibers, $c_{\mathrm{f}} = 0.15$ and (b) $c_{\mathrm{f}} = 0.25$.

## 6 Conclusions

An analytical solution is derived to compute the fluctuations of the local field quantities in random microstructures of fibrous composites. Specifically, an expression is derived for the case of an ellipsoidal-shaped fibers using a singular approximation. The solutions are applicable to two phase composite having isotropic fiber reinforcements in an isotropic matrix with perfectly bonded interface. The fluctuation of stress invariants obtained analytically matches well with solutions obtained semi-analytically. However, it shows minor deviations ranging between 1 - 3 MPa when compared with full-field simulations of random microstructures of fibrous composites. Further it is observed that the volume fraction of fibers significantly affects local field fluctuations whereas the aspect ratio has a negligible impact.

From full-field solutions, it is noticed that the deformation field is nearly homogenous in the particulate phase compared to the matrix phase, which is consistent with the results from the semi-analytical solution. However, because of the high stiffness of the fibers, the stress field fluctuates strongly despite the small fluctuations in the deformation field, which is not captured by the mean-field approaches.

### Acknowledgements

The research documented in this full paper has been funded by the German Research Foundation (DFG) within the International Research Training Group “Integrated engineering of continuous-discontinuous long-fiber reinforced polymer structures” (GRK 2078/2). The support by the German Research Foundation (DFG) is gratefully acknowledged.

# A. Appendix

## A.1 Differentiation of an Inverse of a Fourth Order Tensor

Consider the derivative of the inverse of any major and minor symmetric fourth order tensor $\mathbb{A}$ with a fourth order tensor $\mathbb{B}$. In index notation it is given as

$$\left(\frac{\partial \mathbb{A}^{-1}}{\partial \mathbb{B}}\right)_{ijklmnop} = -A^{-1}_{ijqr} \frac{\partial A_{qrst}}{\partial B_{mnop}} A^{-1}_{stkl}.$$

In tensor symbolic notation, it is given as

$$\frac{\partial \mathbb{A}^{-1}}{\partial \mathbb{B}} = -(\mathbb{A}^{-1})^{\times 2} \frac{\partial \mathbb{A}}{\partial \mathbb{B}}.$$

If $\mathbb{A} = \mathbb{B}$, derivative of any general fourth order tensor with itself would result as

$$\frac{\partial A_{ijkl}}{\partial A_{mnop}} = \delta_{im}\delta_{jn}\delta_{ko}\delta_{lp}.$$

If $\mathbb{A}$ is considered to be major symmetric, differentiation of a fourth order tensor with itself would result as

$$\frac{1}{2}\frac{\partial (A_{ijkl} + A_{klij})}{\partial A_{mnop}} = \frac{1}{2}(\delta_{im}\delta_{jn}\delta_{ko}\delta_{lp} + \delta_{km}\delta_{ln}\delta_{io}\delta_{jp}).$$

In the case when $\mathbb{A}$ exhibits both major and minor symmetry for e.g., like elastic stiffness tensor, derivative of this tensor with itself would result in an eighth order identity tensor which is defined as

$$\begin{aligned}
&\frac{1}{8}\frac{\partial (A_{ijkl} + A_{jikl} + A_{ijlk} + A_{jilk} + A_{klij} + A_{lkij} + A_{klji} + A_{lkji})}{\partial A_{mnop}} \\
&= \frac{1}{8}(\delta_{im}\delta_{jn}\delta_{ko}\delta_{lp} + \delta_{jm}\delta_{in}\delta_{ko}\delta_{lp} + \delta_{im}\delta_{jn}\delta_{lo}\delta_{kp} + \delta_{jm}\delta_{in}\delta_{lo}\delta_{kp} \\
&+ \delta_{km}\delta_{ln}\delta_{io}\delta_{jp} + \delta_{lm}\delta_{kn}\delta_{io}\delta_{jp} + \delta_{km}\delta_{ln}\delta_{jo}\delta_{ip} + \delta_{lm}\delta_{kn}\delta_{jo}\delta_{ip}) = I^{8\mathrm{s}}_{ijklmnop} \\
&= (\boldsymbol{\mathcal{J}}^{\mathrm{s}})_{ijklmnop}.
\end{aligned}$$

Hence the derivative of inverse of a fourth order tensor with major and minor symmetric properties in index notation is given as

$$\left(\frac{\partial \mathbb{A}^{-1}}{\partial \mathbb{A}}\right)_{ijklmnop} = -A^{-1}_{ijqr} I^{8}_{ijklmnop} A^{-1}_{stkl}.$$

In symbolic notation

$$\frac{\partial \mathbb{A}^{-1}}{\partial \mathbb{A}} = -(\mathbb{A}^{-1})^{\times 2} \boldsymbol{\mathcal{J}}^{\mathrm{s}}.$$

### A.2 Simplification of the Integral Term $I_{mn}$

Consider the integral term obtained in Eq. (25)

$$I_{mn} = \int_0^{\pi} f(\theta)(\sin\theta)^{2m+1}(\cos\theta)^{2n}\ \mathrm{d}\theta.$$

where $f(\theta) = (1+\gamma\sin^2\theta)^{-3/2}$ and $\gamma = (\frac{1}{a^2} - 1)$. The integral term can be rewritten as

$$I_{mn} = 2\int_0^{\frac{\pi}{2}} (\sin\theta)^{2m+1}(\cos\theta)^{2n}(1+\gamma\sin^2\theta)^{-\frac{3}{2}}\ \mathrm{d}\theta.$$

Let $u = \sin^2\theta$ and by substitution

$$I_{mn} = \int_0^1 u^m(1-u)^n(1-u)^{-0.5}(1+\gamma u)^{-\frac{3}{2}}\ \mathrm{d}u = \int_0^1 p_{mn}(u)g(u)\ \mathrm{d}u.$$

By using the binomial theorem on $p_{mn}$

$$I_{mn} = \sum_{k=0}^{n} \binom{n}{k} (-1)^k I_{(m+k)0}.$$

Using integration by parts,

$$I_{m0} = \int_0^1 p_{m0} g \mathrm{d}u = [p_{mo}G]_0^1 - \int_0^1 p'_{m0} G \mathrm{d}u = 0 + \frac{2m}{\gamma+1}\int_0^1 p_{(m-1)0}\, g(1-u)(1+\gamma u)\mathrm{d}u$$
$$= \frac{2m}{\gamma+1}\left(\gamma I_{m1} + I_{(m-1)1}\right) = \frac{2m}{\gamma+1}\Big(\gamma(I_{m0} - I_{(m+1)0}) + (I_{(m-1)0} - I_{m0})\Big)$$

$$I_{(m+1)0} = \frac{1}{\gamma}\left(\left(\gamma - 1 - \frac{\gamma+1}{2m}\right) I_{m0} + I_{(m-1)0}\right).$$

### A.3 Definition of Orthonormal Basis Tensors $\mathbb{B}_i$

The sixth order tensors $\mathfrak{J}_1$ and $\mathfrak{J}_2$ is given as (Krause and Böhlke, 2024)

$[\mathfrak{J}_1]_{ijklmn} = [\boldsymbol{I}]_{ij}[\mathbb{P}_2]_{klmn} + [\boldsymbol{I}]_{kl}[\mathbb{P}_2]_{ijmn}$

$[\mathfrak{J}_2]_{abcdef} = [\mathbb{P}_2]_{abij}[\mathbb{P}_2]_{cdkl}([\boldsymbol{I}]_{ik}[\mathbb{P}_2]_{jlef} + [\boldsymbol{I}]_{jl}[\mathbb{P}_2]_{ikef} + [\boldsymbol{I}]_{il}[\mathbb{P}_2]_{jkef} + [\boldsymbol{I}]_{jk}[\mathbb{P}_2]_{ilef})$

$\mathbb{B}_3 = \mathfrak{J}_1[\boldsymbol{D}_1^2] = \mathfrak{J}_1\left[\frac{1}{\sqrt{2}}(\boldsymbol{e}_1 \otimes \boldsymbol{e}_1 - \boldsymbol{e}_2 \otimes \boldsymbol{e}_2)\right]$

$\mathbb{B}_4 = \mathfrak{J}_1[\boldsymbol{D}_2^2] = \mathfrak{J}_1[\sqrt{2}(\boldsymbol{e}_1 \otimes^{\mathrm{s}} \boldsymbol{e}_2)]$

$\mathbb{B}_5 = \mathfrak{J}_1[\boldsymbol{D}_3^2] = \mathfrak{J}_1[\sqrt{2}(\boldsymbol{e}_1 \otimes^{\mathrm{s}} \boldsymbol{e}_3)]$

$\mathbb{B}_6 = \mathfrak{J}_1[\boldsymbol{D}_4^2] = \mathfrak{J}_1[\sqrt{2}(\boldsymbol{e}_2 \otimes^{\mathrm{s}} \boldsymbol{e}_3)]$

$\mathbb{B}_7 = \mathfrak{J}_1[\boldsymbol{D}_5^2] = \mathfrak{J}_1\left[\frac{1}{\sqrt{6}}(3\boldsymbol{e}_3 \otimes \boldsymbol{e}_3 - \boldsymbol{I})\right]$

$\mathbb{B}_8 = \mathfrak{J}_2[\boldsymbol{D}_1^2]$

$\mathbb{B}_9 = \mathfrak{J}_2[\boldsymbol{D}_2^2]$

$\mathbb{B}_{10} = \mathfrak{J}_2[\boldsymbol{D}_3^2]$

$\mathbb{B}_{11} = \mathfrak{J}_2[\boldsymbol{D}_4^2]$

$\mathbb{B}_{12} = \mathfrak{J}_2[\boldsymbol{D}_5^2]$

The fourth-order deviatoric basis tensors $\mathbb{D}_i^4$ can be calculated using the Clebsch-Gordan tensor of appropriate order $\mathbb{c}^{(224)}$, using the formula (Krause and Böhlke, 2024)

$$\mathbb{D}_i^4 = \mathbb{c}_{jki}^{(224)} \boldsymbol{D}_j^2 \otimes \boldsymbol{D}_k^2.$$

The basis tensors $\mathbb{B}_{13}$ to $\mathbb{B}_{21}$ are given as

$$\mathbb{B}_{13} = \mathbb{D}_1^4 = \frac{1}{\sqrt{2}}(\boldsymbol{D}_1^2 \otimes \boldsymbol{D}_1^2 - \boldsymbol{D}_2^2 \otimes \boldsymbol{D}_2^2),$$

$$\mathbb{B}_{14} = \mathbb{D}_2^4 = \frac{1}{\sqrt{2}}(\boldsymbol{D}_1^2 \otimes \boldsymbol{D}_2^2 - \boldsymbol{D}_2^2 \otimes \boldsymbol{D}_1^2),$$

$$\mathbb{B}_{15} = \mathbb{D}_3^4 = \frac{1}{2}(\boldsymbol{D}_1^2 \otimes \boldsymbol{D}_3^2 - \boldsymbol{D}_2^2 \otimes \boldsymbol{D}_4^2 + \boldsymbol{D}_3^2 \otimes \boldsymbol{D}_1^2 - \boldsymbol{D}_4^2 \otimes \boldsymbol{D}_2^2),$$

$$\mathbb{B}_{16} = \mathbb{D}_4^4 = \frac{1}{2}(\boldsymbol{D}_1^2 \otimes \boldsymbol{D}_4^2 + \boldsymbol{D}_2^2 \otimes \boldsymbol{D}_3^2 + \boldsymbol{D}_3^2 \otimes \boldsymbol{D}_2^2 + \boldsymbol{D}_4^2 \otimes \boldsymbol{D}_1^2),$$

$$\mathbb{B}_{17} = \mathbb{D}_5^4 = \frac{1}{\sqrt{14}}(\sqrt{3}\boldsymbol{D}_1^2 \otimes \boldsymbol{D}_5^2 + 2\boldsymbol{D}_3^2 \otimes \boldsymbol{D}_3^2 - 2\boldsymbol{D}_4^2 \otimes \boldsymbol{D}_4^2 + \sqrt{3}\boldsymbol{D}_5^2 \otimes \boldsymbol{D}_1^2),$$

$$\mathbb{B}_{18} = \mathbb{D}_6^4 = \frac{1}{\sqrt{14}}(\sqrt{3}\boldsymbol{D}_2^2 \otimes \boldsymbol{D}_5^2 + 2\boldsymbol{D}_3^2 \otimes \boldsymbol{D}_4^2 + 2\boldsymbol{D}_4^2 \otimes \boldsymbol{D}_3^2 + \sqrt{3}\boldsymbol{D}_5^2 \otimes \boldsymbol{D}_2^2),$$

$$\begin{aligned}\mathbb{B}_{19} = \mathbb{D}_7^4 = \frac{1}{2\sqrt{7}}(&-\boldsymbol{D}_1^2 \otimes \boldsymbol{D}_3^2 - \boldsymbol{D}_2^2 \otimes \boldsymbol{D}_4^2 - \boldsymbol{D}_3^2 \otimes \boldsymbol{D}_1^2 + 2\sqrt{3}\boldsymbol{D}_3^2 \otimes \boldsymbol{D}_5^2 - \boldsymbol{D}_4^2 \otimes \boldsymbol{D}_2^2 \\ &+ 2\sqrt{3}\boldsymbol{D}_5^2 \otimes \boldsymbol{D}_3^2),\end{aligned}$$

$$\begin{aligned}\mathbb{B}_{20} = \mathbb{D}_8^4 = \frac{1}{2\sqrt{7}}(&\boldsymbol{D}_1^2 \otimes \boldsymbol{D}_4^2 - \boldsymbol{D}_2^2 \otimes \boldsymbol{D}_3^2 - \boldsymbol{D}_3^2 \otimes \boldsymbol{D}_2^2 + \boldsymbol{D}_4^2 \otimes \boldsymbol{D}_1^2 + 2\sqrt{3}\boldsymbol{D}_4^2 \otimes \boldsymbol{D}_5^2 \\ &+ 2\sqrt{3}\boldsymbol{D}_5^2 \otimes \boldsymbol{D}_4^2),\end{aligned}$$

$$\mathbb{B}_{21} = \mathbb{D}_9^4 = \frac{1}{\sqrt{70}}(\boldsymbol{D}_1^2 \otimes \boldsymbol{D}_1^2 + \boldsymbol{D}_2^2 \otimes \boldsymbol{D}_2^2 - 4\boldsymbol{D}_3^2 \otimes \boldsymbol{D}_3^2 - 4\boldsymbol{D}_4^2 \otimes \boldsymbol{D}_4^2 + 6\boldsymbol{D}_5^2 \otimes \boldsymbol{D}_5^2).$$

## References

Albiez, J., Erdle, H., Weygand, D., Böhlke, T., 2019. A gradient plasticity creep model accounting for slip transfer/activation at interfaces evaluated for the intermetallic NiAl-9Mo. Int. J. Plast. 113, 291–311. https://doi.org/10.1016/j.ijplas.2018.10.006

Asp, L.E., Berglund, L.A., Talreja, R., 1996. A criterion for crack initiation in glassy polymers subjected to a composite-like stress state. Compos. Sci. Technol. 56, 1291–1301. https://doi.org/10.1016/S0266-3538(96)00090-5

Badulescu, C., Lahellec, N., Suquet, P., 2015. Field statistics in linear viscoelastic composites and polycrystals. Eur. J. Mech. A/Solids 49, 329–344. https://doi.org/10.1016/j.euromechsol.2014.07.012

Bobeth, M., Diener, G., 1987. Static elastic and thermoelastic field fluctuations in multiphase composites. J. Mech. Phys. Solids 35, 137–149. https://doi.org/10.1016/0022-5096(87)90033-0

Bobeth, M., Diener, G., 1986. Field fluctuations in multicomponent mixtures. J. Mech. Phys. Solids 34, 1–17. https://doi.org/10.1016/0022-5096(86)90002-5

Brenner, R., Castelnau, O., Badea, L., 2004. Mechanical Field Fluctuations in Polycrystals Estimated by Homogenization Techniques. Source Proc. Math. Phys. Eng. Sci. 460, 3589–3612. https://doi.org/10.1098/rspa.2004.1278

Castañeda, P.P., Willis, J.R., 1995. The effect of spatial distribution on the effective behavior of composite materials and cracked media. J. Mech. Phys. Solids 43, 1919–1951. https://doi.org/10.1016/0022-5096(95)00058-Q

Das, S., Ponte Castañeda, P., 2021. Field statistics in linearized elastic and viscous composites and polycrystals. Int. J. Solids Struct. 224, 111030.

https://doi.org/10.1016/j.ijsolstr.2021.03.017

Elnekhaily, S.A., Talreja, R., 2023. Effects of micro voids on the early stage of transverse crack formation in unidirectional composites. Compos. Part A Appl. Sci. Manuf. 167, 107457. https://doi.org/10.1016/j.compositesa.2023.107457

Elnekhaily, S.A., Talreja, R., 2019. Effect of axial shear and transverse tension on early failure events in unidirectional polymer matrix composites. Compos. Part A Appl. Sci. Manuf. 119, 275–282. https://doi.org/10.1016/j.compositesa.2019.01.031

Estevez, R., Tijssens, M.G.A., Van Der Giessen, E., 2000. Modeling of the competition between shear yielding and crazing in glassy polymers. J. Mech. Phys. Solids 48, 2585–2617. https://doi.org/10.1016/S0022-5096(00)00016-8

Fokin, A.G., 1974. Solution of statistical problems in elasticity theory in the singular approximation. J. Appl. Mech. Tech. Phys. 13, 85–89. https://doi.org/10.1007/BF00852360

Forte, S., Vianello, M., 1996. Symmetry classes for elasticity tensors. J. Elast. 43, 81–108. https://doi.org/10.1007/BF00042505

Gagarinov, P., Kurzhanskiy, A.A., 2014. Ellipsoidal Toolbox Release 2.0.1.

Görthofer, J., Meyer, N., Pallicity, T.D., Schöttl, L., Trauth, A., Schemmann, M., Hohberg, M., Pinter, P., Elsner, P., Henning, F., Hrymak, A., Seelig, T., Weidenmann, K., Kärger, L., Böhlke, T., 2019. Virtual process chain of sheet molding compound: Development, validation and perspectives. Compos. Part B Eng. 169. https://doi.org/10.1016/j.compositesb.2019.04.001

Hashin, Z., Shtrikman, S., 1962. A variational approach to the theory of the elastic behaviour of polycrystals. J. Mech. Phys. Solids 10, 343–352.

https://doi.org/10.1016/0022-5096(62)90005-4

Hessman, P.A., Welschinger, F., Hornberger, K., Böhlke, T., 2021. On mean field homogenization schemes for short fiber reinforced composites: Unified formulation, application and benchmark. Int. J. Solids Struct. 230–231, 111141. https://doi.org/10.1016/j.ijsolstr.2021.111141

Hu, G.K., Weng, G.J., 2000. Connections between the double-inclusion model and the Ponte Castaneda-Willis, Mori-Tanaka, and Kuster-Toksoz models. Mech. Mater. 32, 495–503. https://doi.org/10.1016/S0167-6636(00)00015-6

Idiart, M.I., Moulinec, H., Ponte Castañeda, P., Suquet, P.M., 2006. Macroscopic behavior and field fluctuations in viscoplastic composites: Second-order estimates versus full-field simulations. J. Mech. Phys. Solids 54, 1029–1063. https://doi.org/10.1016/j.jmps.2005.11.004

Jöchen, K., 2013. Homogenization of the Linear and Non-linear Mechanical Behavior of Polycrystals. KIT Sci. Publ. Karlsruhe. Doctoral Thesis, Institute of Engineering Mechanics (ITM), Department of Mechanical Engineering, Karlsruhe Institute of Technology (KIT), Karlsruhe. https://doi.org/10.5445/KSP/1000032289

Kehrer, L., 2019. Thermomechanical mean-field modeling and experimental characterization of long fiber-reinforced sheet molding compound composites. Doctoral Thesis, Institute of Engineering Mechanics (ITM), Department of Mechanical Engineering, Karlsruhe Institute of Technology (KIT), Karlsruhe.

Kowalczyk-Gajewska, K., Berbenni, S., Mercier, S., 2024. An additive Mori–Tanaka scheme for elastic–viscoplastic composites based on a modified tangent linearization. Mech. Mater. 200, 105191. https://doi.org/10.1016/j.mechmat.2024.105191

Krause, M., Böhlke, T., 2024. Tensorial harmonic bases of arbitrary order with applications in elasticity, elastoviscoplasticity and texture-based modeling. Math. Mech. Solids. https://doi.org/10.1177/10812865241247519

Krause, M., Pallicity, T.D., Böhlke, T., 2023. Exact second moments of strain for composites with isotropic phases. Eur. J. Mech. - A/Solids 97, 104806. https://doi.org/10.1016/j.euromechsol.2022.104806

Lahellec, N., Suquet, P., 2007a. Effective behavior of linear viscoelastic composites: A time-integration approach. Int. J. Solids Struct. 44, 507–529. https://doi.org/10.1016/j.ijsolstr.2006.04.038

Lahellec, N., Suquet, P., 2007b. On the effective behavior of nonlinear inelastic composites: I. Incremental variational principles. J. Mech. Phys. Solids 55, 1932–1963. https://doi.org/10.1016/J.JMPS.2007.02.003

Lahellec, N., Suquet, P., 2007c. On the effective behavior of nonlinear inelastic composites: II: A second-order procedure. J. Mech. Phys. Solids 55, 1964–1992. https://doi.org/10.1016/J.JMPS.2007.02.004

Levin, V.M., 1967. Thermal expansion coefficients of heterogeneous materials. Mech. Solids 2, 58–61.

Lopez-Pamies, O., Castañeda, P.P., 2006. On the overall behavior, microstructure evolution, and macroscopic stability in reinforced rubbers at large deformations: I - Theory. J. Mech. Phys. Solids 54, 807–830. https://doi.org/10.1016/j.jmps.2005.10.006

Mori, T., Tanaka, K., 1973. Average stress in matrix and average elastic energy of materials with misfitting inclusions. Acta Metall. 21, 571–574.

Moulinec, H., Suquet, P., 1998. A numerical method for computing the overall response of nonlinear composites with complex microstructure. Comput. Methods Appl. Mech. Eng. 157, 69–94. https://doi.org/10.1016/S0045-7825(97)00218-1

Pallicity, T.D., Böhlke, T., 2021. Effective viscoelastic behavior of polymer composites with regular periodic microstructures. Int. J. Solids Struct. 216, 167–181. https://doi.org/10.1016/j.ijsolstr.2021.01.016

Ponte Castañeda, P., 2002. Second-order homogenization estimates for nonlinear composites incorporating field fluctuations: I—theory. J. Mech. Phys. Solids 50, 737–757. https://doi.org/10.1016/S0022-5096(01)00099-0

Schneider, M., 2021. A review of nonlinear FFT-based computational homogenization methods. Acta Mech. 232, 2051–2100. https://doi.org/10.1007/s00707-021-02962-1

Schöberl, J., 1997. NETGEN An advancing front 2D/3D-mesh generator based on abstract rules. Comput. Vis. Sci. 1, 41–52. https://doi.org/10.1007/s007910050004

Talreja, R., 2006. Multi-scale modeling in damage mechanics of composite materials. J. Mater. Sci. 41, 6800–6812. https://doi.org/10.1007/s10853-006-0210-9

Torquato, S., 2002. Random Heterogeneous Materials, 16th ed, Interdisciplinary Applied Mathematics. Springer New York, New York, NY. https://doi.org/10.1007/978-1-4757-6355-3

Trauth, A., Weidenmann, K.A., 2018. Continuous-discontinuous sheet moulding compounds – Effect of hybridisation on mechanical material properties. Compos. Struct. 202, 1087–1098. https://doi.org/10.1016/j.compstruct.2018.05.048

Walpole, L.J., 1966. On bounds for the overall elastic moduli of inhomogeneous

systems-I. J. Mech. Phys. Solids 14, 151–162. https://doi.org/10.1016/0022-5096(66)90035-4

Willis, J.R., 1977. Bounds and self-consistent estimates for the overall properties of anisotropic composites. J. Mech. Phys. Solids 25, 185–202. https://doi.org/10.1016/0022-5096(77)90022-9

Willot, F., Brenner, R., Trumel, H., 2020. Elastostatic field distributions in polycrystals and cracked media. Philos. Mag. 100, 661–687. https://doi.org/10.1080/14786435.2019.1699669

Zheng, Q.-S., Spencer, A.J.M., 1993. On the canonical representations for kronecker powers of orthogonal tensors with application to material symmetry problems. Int. J. Eng. Sci. 31, 617–635. https://doi.org/10.1016/0020-7225(93)90054-X